\documentclass[aps,prd,a4paper,showpacs,12pt]{article}
\usepackage[margin=1.0in]{geometry}
\usepackage{multicol}
\usepackage{flushend}
\usepackage[toc,page]{appendix}
\usepackage{subfigure}
\usepackage {graphicx}
\usepackage{color}
\usepackage{xcolor}
\usepackage{threeparttable}
\usepackage{amsfonts}
\usepackage{times}
\usepackage{setspace}
\usepackage{balance}
\usepackage{lastpage}
\usepackage{amsthm}
\usepackage[tbtags]{amsmath}
\usepackage{nomencl}
\usepackage{pstricks}
\usepackage{pstricks-add}
\usepackage{pst-plot,pstricks-add}

\usepackage{amsmath,amsfonts,amssymb,simplewick}
\usepackage{float}

\usepackage{amsmath,amssymb,amsfonts,epsfig,graphicx,euscript}%
\usepackage{hyperref} 
\hypersetup{bookmarks=true, bookmarksnumbered=true,linktoc=page, pdfstartview={FitH}, 
	colorlinks=true, 
	citecolor=red, 
	filecolor=blue, 
	linkcolor=blue, 
	urlcolor=blue}
\usepackage{amsmath}
\usepackage{amsfonts}
\usepackage{graphicx}
\usepackage{caption}
\usepackage{float}
\usepackage[all]{hypcap}
\numberwithin{equation}{section}
\usepackage{cite}
\usepackage{calc}
\newlength{\spacer}
\newsavebox{\mybox}

\newcommand{\bse}{\begin{subequations}}
	\newcommand{\ese}{\end{subequations}}
\newcommand{\be}{\begin{equation}}
	\newcommand{\ee}{\end{equation}}
\newcommand{\bea}{\begin{eqnarray}}
	\newcommand{\eea}{\end{eqnarray}}
\newcommand{\ba}{\begin{array}}
	\newcommand{\ea}{\end{array}}

\renewcommand{\thefootnote}{\fnsymbol{footnote}}
\begin{document}
	\begin{center}
		{\large{\textbf{Global charge conservation in the symmetric phase of the early Universe}}} 
		\vspace*{1.5cm}
		\begin{center}
			{\bf S. Abbaslu\footnote{s$_{-}$abbasluo@sbu.ac.ir}$^{1,2}$, M. Abdolhoseini\footnote{m4.abdolhoseini1395@gmail.com}$^2$, P. E. Moghaddam \footnote{poupak$_{-}$75@yahoo.com}$^2$, and S. S. Gousheh\footnote{ss-gousheh@sbu.ac.ir}$^2$}\\
			\vspace*{0.5cm}
			{\it{$^1$ School of Physics, Institute for Research in Fundamental Sciences (IPM), PO Box 19395-5531, Tehran, Iran
            \\$^2$ Department of Physics, Shahid Beheshti University, Tehran, Iran}}\\
			
			\vspace*{1cm}
		\end{center}
	\end{center}
	\begin{center}
		\today
	\end{center}
	\renewcommand*{\thefootnote}{\arabic{footnote}}
	\setcounter{footnote}{0}
	
	\date{\today}
	\textbf{Abstract:} In the Standard Model at high temperatures, anomalous effects contribute to the violation of baryon number ($B$) and lepton number ($L$), separately, while $B-L$ remains conserved. There are also corresponding changes in the helicity of the hypermagnetic field ($h_B$) and the Chern-Simons numbers of the non-Abelian gauge fields ($N_{\rm CS,w}$ and $N_{\rm CS,s}$).  In this study, we investigate a baryogenesis process in the symmetric phase of the early Universe by taking into account the Abelian and non-Abelian anomalous effects as well as the perturbative chirality-flip processes of all fermions.  We calculate the time evolution of all relevant physical quantities, including the asymmetries of all fermions and the Higgs, as well as $h_B$ and $N_{\rm CS,w}$. We present a method to compute the latter, for which it is crucial to consider the minute departure from equilibrium of the sphaleron processes. We then verify explicitly the conservation of a global charge, involving the total matter-antimatter asymmetry $B+L$, the hypermagnetic helicity $h_B$ and $N_{\rm CS,w}$, the existence of which had been inferred earlier. In particular we show that, in the scenario that we study, an initial $h_B$ decays mostly to $N_{\rm CS,w}$, with only $10^{-3}$ conversion ratio into $B+L$ asymmetry.
	
	

\section{Introduction}

The baryon asymmetry of the Universe remains a longstanding problem in the realms of cosmology and particle physics. The measured baryon asymmetry of the Universe is of the order of $\eta_B\sim10^{-10}$ \cite{baryon3,WMAP-2010qai}. The idea of baryogenesis can be traced back to 1967 when Sakharov proposed that the baryon asymmetry is not a fundamental property of the Universe from the beginning but rather it could be generated through particle physics processes at a later stage \cite{Sakharov1}. This concept has also found support from the inflationary scenario, as inflation is believed to have diluted any pre-existing asymmetry that may have been present initially. 

On the other hand, observations indicate that our Universe is magnetized on various scales \cite{1,2}. Lower bounds and signatures of coherent intergalactic magnetic fields of the order $B \sim 10^{-15}\text{ G}$ have been inferred from blazar observations \cite{magnetic1,magnetic2,magnetic3,magnetic4,magnetic5}. While early claims of detection (such as Ref.~\cite{magnetic3}) have been subject to ongoing debate regarding systematic uncertainties in Fermi-LAT data \cite{Neronov:2010bi,Fermi-LAT:2013mql,AlvesBatista:2021sln}, the observational status of intergalactic magnetic fields remains an active area of study.
 Generally there are two major approaches for studying the generation and evolution of coherent magnetic fields, namely the astrophysical models \cite{Durrer-2013pga,Naoz-2013wla} and cosmological models \cite{mf1,mf2,mf4,mf5,mf6,mf7}. Recent observations \cite{Taylor-2011bn,huan,Vovk}, combined with the ubiquitous presence of large-scale magnetic fields throughout the Universe, strengthen the hypothesis that these magnetic fields originated in the early Universe, consistent with the predictions of the cosmological models \cite{Kandus-2010nw}.

In the standard model, lepton and  baryon numbers are conserved at low energies. However, as temperature rises above $100$ GeV, the symmetry of the $ {\rm SU(2)_L}\times{\rm U(1)}_Y$ gauge group is restored, leading to the possibility of violation of lepton and quark current conservations due to the effects of triangle anomalies \cite{a1,a2,a3}. Moreover, there are anomalous transport phenomena, {\it i.e.},
the chiral vortical effect (CVE) and the chiral magnetic effect (CME) \cite{Vilenkin:1978hb,av1,avi1}, that play significant roles in particle physics and cosmology, particularly in the early Universe \cite{kharz}. The CVE refers to the generation of an electric current parallel to the vorticity, while the CME involves the generation of an electric current parallel to the magnetic field in a chiral plasma. These effects have been studied extensively due to their importance in understanding the dynamics of systems with chiral fermions  \cite{Giovannini-1997eg,Giovannini-2013oga,Giovannini-2016whv}. To account for these anomalous effects, the ordinary magnetohydrodynamic (MHD) equations are generalized to the framework of anomalous magnetohydrodynamics (AMHD) \cite{Giovannini-1997eg,Giovannini-2013oga,Giovannini-2016whv}. Recently, there have been numerous studies which use the Abelian and non-Abelian anomalies along with AMHD equations to present scenarios for the production of matter-antimatter asymmetries and long-range hypermagnetic fields \cite{Giovannini-1997eg,Giovannini-2013oga,Giovannini-2016whv,s2,s3,s4}. As we shall show explicitly, there are corresponding changes in the vacuum sectors of the non-Abelian gauge fields, which play a prominent role. After the electroweak phase transition, while the baryon asymmetry remains constant, these hypermagnetic fields predominantly transform into  Maxwellian magnetic fields \cite{baryon-5}.  While our present analysis focuses on the symmetric phase up to the electroweak crossover, the subsequent dynamics during the crossover itself are known to play a key role in setting the final pre-BBN baryon asymmetry \cite{baryon-5, Hamada:2025cwu}. Given that the precise quantitative impact of this phase transition remains an active area of investigation with ongoing refinements \cite{Boyer:2025jno,Hamada:2025ooy}, extending our full numerical framework through the crossover regime constitutes an important direction for future work.
	
	 	
The electroweak sector has an infinite number of degenerate and topologically distinct vacua \cite{Jackiw-1976pf,Callan-1976je,Manton-1983nd,Klinkhamer-1984di}, separated by energy barriers and characterized by integer values of the Chern-Simons number,
		\begin{equation}\label{eqr}
		\begin{split}
			 \frac{ d N_{\rm CS,w}(t)}{dt}=\frac{g^2}{16\pi^2}\int d^3 x {\rm Tr} [W_{\mu\nu}\tilde{W}^{ \mu\nu}],
		\end{split}
	\end{equation} 
where $g$ is the weak coupling constant, $W_{\mu\nu}$ denotes the field strength tensor of the $\rm SU(2)_L$ gauge group, and $\tilde{W}^{ \mu\nu}$ represents the dual of $W^{ \mu\nu}$ defined as $\tilde{W}^{\mu\nu}=\frac{1}{2R^3}\epsilon^{\mu\nu\rho\sigma}W_{\rho\sigma}$,  with the totally anti-symmetric Levi-Civita tensor density specified by $\epsilon^{0123}=-\epsilon_{0123}=1$ \cite{Kuzmin-1985mm,Arnold-987mh,Rubakov-1996vz}.
This Chern-Simons number serves as a topological charge for the $\rm SU(2)_L$ gauge field configurations.  The $\rm{SU(2)_L}$ gauge fields couple to the left-handed quarks and leptons. 
When the system undergoes a transition from one vacuum with a Chern-Simons number $\rm N_{\rm CS,w}$ to a neighboring vacuum with $\rm N_{\rm CS,w}\pm1$, profound consequences arise \cite{Kuzmin-1985mm,Arnold-987mh,Rubakov-1996vz}. These transitions, for example, result in the creation or annihilation of left-handed quarks and leptons \cite{Kuzmin-1985mm}.	
In the electroweak sector, the barrier between adjacent vacua in the broken phase is approximately $10\ \rm TeV$ and is proportional to $v/g$, where $v$ is the Higgs vacuum expectation value \cite{Manton-1983nd}. Therefore, in this phase, the instanton processes which lead to tunneling through the barrier are highly suppressed, as are the sphaleron processes which represent classical solutions traversing the barrier. At higher temperatures of the symmetric phase, the barriers remaining are only due to the finite temperature effective potential, which can be traversed by the sphaleron processes with relative ease and at the rate $\Gamma_{\rm w} \simeq 25\alpha_{\rm w}^5T$.

The $\rm{SU(3)}$ gauge fields couple to both left-handed and right-handed quarks equally. 
The strong sphaleron processes refer to the vacuum-to-vacuum transitions in the $\rm{SU(3)}$ sector. 
It is important to note that while the strong sphaleron processes can lead to violation of chiral quark numbers, they preserve the baryon number. 
The reaction rate for these processes is approximately $\Gamma_{\rm s}\simeq 100\alpha_{\rm s}^5 T$, where $\alpha_{\rm s}$ represents the $\rm{SU(3)}$ fine structure constant \cite{Moore-1,Moore-2010jd}. These processes become effective when their reaction rate exceeds the expansion rate of the Universe. Typically, this occurs at temperatures below $T\sim 10^{15}\,\rm{GeV}$.

Unlike the non-Abelian cases, the ${\rm U}(1)_Y$ gauge field couples chirally to all quarks and leptons, and there are no sphaleron-like processes due to its trivial topology. However, quark and lepton number violations still occur, due to the corresponding triangle anomaly, through the time variation of background hypermagnetic field helicity \cite{Giovannini-1997eg}. This specific scenario has been extensively studied to explain the observed baryon asymmetry of the Universe \cite{ms1,baryon-1,Dvornikov:2022gkf,baryon-2,baryon-3,baryon-4,baryon-5,baryon-6,baryon-7,baryon-8}.

In this paper we study a model for the generation of matter-antimatter asymmetry in the presence of helical hypermagnetic fields in the symmetric phase of the early Universe, starting at $T=10$ TeV and ending at the onset of the electroweak phase transition, which we assume to be at $T=100$ GeV.  We include the Abelian and non-Abelian anomalous effects
as well as the perturbative chirality-flip processes and the CME \cite{avi1}, while assuming, for simplicity, that there is no fluid vorticity in the electroweak plasma, and hence no CVE\footnote{This assumption is valid for our case where the initial hypermagnetic fields is sufficiently large, ensuring that the chiral vortical effect (CVE) remains negligible compared to the chiral magnetic effect (CME). Moreover, for helical fields and fluids, the advection term has no impact on the hypermagnetic field evolution.}.
The anomalous processes contribute to the violation of baryon number ($B$) and lepton number ($L$), separately, while $B-L$ remains conserved. There are also corresponding changes in the helicity of the Abelian gauge field ($h_B$) and the Chern-Simons numbers of the non-Abelian gauge fields ($N_{\rm CS,w}$ and $N_{\rm CS,s}$). Besides $B-L$, the existence of another conserved global charge involving $B$, $L$, $h_B$, and $N_{\rm CS,w}$ has been inferred \cite{h1,baryon-2}.\footnote{ In particular, \cite{baryon-2}  has focused on determining the final baryon asymmetry through the electroweak crossover, while our work emphasizes the dynamical conservation of the total Matter-Gauge charge and provides a quantitative tracking of how the decay of hypermagnetic helicity is divided up between the fermion asymmetries and the weak Chern--Simons number in the symmetric phase.}  To investigate these, we calculate the time evolution of all relevant physical quantities, including the asymmetries of all leptons and quarks, $h_B$ and $N_{\rm CS,w}$. To compute the latter, we introduce a method which keeps track of minute deviation of weak sphaleron processes from equilibrium. We then explicitly verify that there is another conserved global charge involving total matter-antimatter asymmetry $B+L$, $h_B$, and $N_{\rm CS,w}$. In our computations, only the strong sphaleron processes are assumed to be in equilibrium, which, at any rate, do not contribute to baryon number violation. In particular, we show that the decay of positive hypermagnetic helicity $h_B$ results in the production of positive $B+L$ and a positive change in $N_{\rm CS,w}$, with approximately 1:1000 production ratio. To be concrete, in the evolution equations of the matter fields that we present in Sec.~\ref{4}, the ${\rm U(1)}_Y$ anomaly appears directly as the rate of change of magnetic helicity, while the $\rm SU(2)_L$ anomaly appears as the sphaleron-mediated interactions. Also, In the temperature range under consideration, all perturbative gauge interactions are fast enough to be included as constraints, while the perturbative Yukawa interaction appear as chirality-flip processes. We then show that when we add up all these evolution equations, the perturbative processes cancel out, and the remaining terms, all originating from the anomaly equations, can be presented as a conservation law.

The rest of this paper is organized as follows: In Sec.\ (\ref{arew1}), we write the  anomaly equations in the electroweak plasma, and determine the globally conserved currents in the absence of perturbative gauge and chirality-flip processes. Section (\ref{4}) is dedicated to the evolution equations for the chiral leptons and baryon asymmetries, considering both perturbative and nonperturbative effects. Subsequently, we present an expression for the  globally conserved charge of the electroweak plasma, which includes the weak sphaleron Chern-Simons number and the hypermagnetic field helicity. In  Sec.\ (\ref{discussion}), the evolution equations are numerically solved, first for a single mode hypermagnetic field and next for a continuous distribution, and the results are presented. In particular, the constancy of the global charge is explicitly demonstrated. Finally, our conclusions are given in Sec.\ (\ref{x5}). Moreover, in App.\ (\ref{ame}) the anomalous Maxwell equations in an expanding Universe are presented, which include the CME. Lastly, App.\ (\ref{APP-B}) details the evolution equations for continuous hypermagnetic field spectra.

\section{Anomaly equations in the electroweak plasma} \label{arew1}

In the presence of anomalous effects, the conservation of global chiral currents of fermions are violated. The anomaly equations for chiral fermionic currents in an expanding Universe  can be expressed in the following compact form \cite{a1,a2,a3,baryon-2}
\begin{equation}\label{er}
		\begin{split}
			\nabla_{\mu} \tilde{j}_{{\rm f}_{r}^{i}}^{\mu}=-\frac{1}{4}(r N_{\rm c}N_{\rm w}Y_{{\rm f}_r}^{2})\frac{g'^{2}}{16\pi^2}Y_{\mu\nu}\tilde{Y}^{\mu\nu}+\frac{1}{2}(N_{\rm c}a_{\rm w})\frac{g^{2}}{16\pi^2}W_{\mu\nu}^{a}\tilde{W}^{a\, \mu\nu}-\frac{1}{2}(r a_{\rm c}N_{\rm w})\frac{g_{\rm s}^{2}}{16\pi^2}G_{\mu\nu}^{A}\tilde{G}^{A\, \mu\nu},
		\end{split}
	\end{equation} 
where $\nabla_{\mu}$ denotes the covariant derivative with respect to the FRW metric, $\tilde{j}^{\mu}_{{\rm f}_{r}^{i}}$ denote the generalized chiral matter currents with chiralities specified by $r=\pm 1$, {\it i.e.}, for singlets ${{\rm f}_{+1}^{i}}={e}_{R}^i, {d}_{R}^{i}, {u}_{R}^{i} $, and for doublets ${{\rm f}_{-1}^{i}}={l}_{L}^i, {q}_{L}^{i} $, with $l_{L}^i=e_{L}^i+\nu_{L}^i$ and $q_{L}^i=u_{L}^i+d_{L}^i$. Here, $i$ is the generation index, $Y_{{\rm f}_r}$  represents the relevant hypercharge, $N_{\rm c}$ indicates the corresponding rank of the non-Abelian $\rm SU(3)$ gauge group\footnote{We assume that, due to the fast $\rm SU(3)$ color interaction at high temperature, all up or down quarks with different colors have the same chemical potential, {\it i.e.,} for each up and down quark generation we have $\mu_{q_{\rm red}}=\mu_{q_{\rm blue}}=\mu_{q_{\rm green}}$.} (3 for quarks and 1 for leptons), $N_{\rm w}$ indicates the corresponding rank of the non-Abelian $\rm SU(2)_L$ gauge group ($N_{\rm w}=2$ for left-handed and $N_{\rm w}=1$ for right-handed fermions), $a_{\rm w}$ is the $\rm SU(2)_L$ factor (1 for left-handed and 0 for right-handed fermions), $a_{\rm c}$ is the $\rm SU(3)$ factor (1 for quarks and 0 for leptons), and $g_{\rm s}$, $g$ and $g'$ are the coupling constants for ${\rm SU}(3)$, $\rm SU(2)_L$ and ${\rm U}_Y(1)$, respectively. Moreover, the field strength tensors $G_{\mu \nu}^{A}$, $W_{\mu \nu}^{a}$ and $Y_{\mu \nu}$  are given by \cite{baryon-2,h1}
\begin{equation}
		\begin{split}
			&G_{\mu\nu}^{A}=\nabla_{\mu}G_{\nu}^{A}-\nabla_{\nu}G_{\mu}^{A}+g_{\rm s} t^{ABC}G_{\mu}^{B}G_{\nu}^{C},\\& W_{\mu\nu}^{a}=\nabla_{\mu}W_{\nu}^{a}-\nabla_{\nu}W_{\mu}^{a}+g f^{abc}W_{\mu}^{b}W_{\nu}^{c},\\& Y_{\mu\nu}=\nabla_{\mu}Y_{\nu}-\nabla_{\nu}Y_{\mu},
		\end{split}
\end{equation}	
where $t^{ABC}$ and $f^{abc}$ are the structure constants of $\rm SU(3)$ and $\rm SU(2)_L$, respectively. The anomaly equations, by definition, do not include the perturbative processes. In the next section we derive the time evolution equations by adding the perturbative processes, comprising of the gauge and Yukawa interactions, to the anomaly equations. The former are taken into account in the form of constraints.
	
The generalized current	$\tilde{j}_{{\rm f}_{r}^{i}}^{\mu}$ for each fermion species is 
\begin{equation}\label{am}
		\begin{split}
			\tilde{j}_{{\rm f}_{r}^{i}}^{\mu}=N_{\rm c} n_{_{r}^{i}}u^{\mu}+\xi_{ {B,{\rm f}_{r}^{i}}}B^{\mu}+\xi_{{\rm v,f}_{r}^{i}}\omega^{\mu}+\sigma_{{\rm f}_{r}^{i}}E^{\mu}.
		\end{split}
\end{equation}
The first term on the right-hand side of the above equation represents the chiral matter currents, in which $n_{{\rm f}_{r}^{i}} $ denotes the difference between number densities of particles and antiparticles, and $u^{\mu}=\gamma\left(1,\vec{v}/R\right)$ is the four-velocity of the plasma normalized such that $u^{\mu}u_{\mu}=1$. Moreover,  $B^{\mu}=(\epsilon^{\mu\nu\rho\sigma}/2)u_{\nu}Y_{\rho\sigma}$ is the hypermagnetic field four-vector, $\omega^{\mu}=(\epsilon^{\mu\nu\rho\sigma})u_{\nu}\partial_{\rho}u_{\sigma}$ is the vorticity four-vector, $E^{\mu}=F^{\mu\nu}u_{\nu}$ is the electric field four-vector,
$\xi_{B,{\rm f}_{r}^{i}}$ denotes the chiral magnetic coefficient, $\xi_{\mathrm {v,{f}_{r}^{i}}}$ represents the chiral vortical coefficient, and $\sigma_{{\rm f}_{r}^{i}}$ denotes the conductivity coefficient, which are given by
\begin{equation}\label{eq17-wy2q3q}
		\begin{split}
			&\xi_{B,{\rm f}_{r}^{i}}=-rN_{\rm c}\frac{g'}{8\pi^{2}}\big[Y_{\rm f} \mu_{{\rm f}_{r}^{i}}\big],\quad\xi_{\mathrm {v,{f}_{r}^{i}}}=rN_{\rm c}\Big[(\frac{\mu_{{\rm f}_{r}^{i}}^2}{8\pi^2}+\frac{T^2}{24})\Big],\quad \sigma_{{\rm f}_{r}^{i}}\sim N_{\rm c}  \frac{T}{\alpha_{{\rm f}_{r}^i}\ln(1/ \alpha_{{\rm f}_{r}^i})}.
		\end{split}
\end{equation}
Here, we have used $\sigma_{{\rm f}_{r}^{i}}\sim \alpha_{{\rm f}_{r}^i} n_{{\rm f}_{r}^{i}}^{\star}\tau_{\rm f}/T$ \footnote{Here, $n_{{\rm f}_{r}^{i}}^{\star}\sim T^3$ is the number density of particles not the asymmetry $n_{{\rm f}_{r}^{i}}\sim \mu T^2$.} for conductivity coefficient, $\tau\sim\big[ {\alpha_{{\rm f}_{r}^i}}^2 \ln(1/ \alpha_{{\rm f}_{r}^i})T\big]^{-1}$ being the characteristic hyperelectric relaxation time \cite{Baym-1997gq,Arnold-2000dr} and $\alpha_{{\rm f}_{r}^i}={Y_{{\rm f}_{r}^i}}^2\alpha_Y$ with the following relevant hypercharges: 
\begin{equation}\label{eqds1}
\begin{split}
Y_{e_{L}}=-1,\ \ Y_{e_{R}}=-2,\ \ Y_{Q}=\frac{1}{3},\ \ Y_{u_{R}}=\frac{4}{3},\ \ Y_{d_{R}}=-\frac{2}{3}.
\end{split}	
\end{equation} 
Henceforth, we will focus on the zero velocity limit in our model for the sake of simplicity.\footnote{	In our current treatment, we neglect the velocity field of the electroweak plasma ($\mathbf{v} \approx 0$), thereby omitting the advection and stretching terms in induction equation, as well as convective electric field term ($\mathbf{v} \times \mathbf{B}$) in Ohm's law. While this assumption isolates the non-trivial interplay between chiral fermion dynamics and hypermagnetic helicity, a complete description of primordial magnetic field evolution generally requires solving the full system of anomalous magnetohydrodynamics (AMHD) equations including plasma turbulence and velocity dynamics \cite{Banerjee:2004df,Brandenburg:2017neh,RoperPol:2025lgc}. Incorporating fluid velocity dynamics into our multi-flavor framework constitutes a promising direction for future work.} In this frame, the temporal component of all four vectors becomes zero, while their spatial components retain their usual form, {\it i.e.,} $X^{\mu}=(0,\vec{X})$. Moreover, this amounts to setting the CVE to zero, as mentioned earlier.

Upon using the definition of the field strength tensors, we can write the  right-hand side of the Eq.\ (\ref{er}) as four-divergence of currents \cite{h1,baryon-2},  
\begin{equation}\label{jbjlq2}
		\begin{split}
			& \nabla_{\mu}\tilde{j}_{{\rm f}_{r}^{i}}^{\mu}=-rN_{\rm c}N_{\rm w}C_{{\rm f}_{r}^{i}}^Y\nabla_{\mu}K_{Y}^{\mu}+ra_{\rm w}N_{\rm c}C_{{\rm f}_{r}^{i}}^{W}\nabla_{\mu}K_{W}^{\mu}-r a_{\rm c}N_{\rm w}C_{{\rm f}_{r}^{i}}^{G}\nabla_{\mu}K_{G}^{\mu},
		\end{split}
\end{equation} 
where $C_{{\rm f}_{r}^{i}}^Y=\frac{Y_{\rm f}^2}{4}\frac{1}{16\pi^2}$, $C_{{\rm f}_{r}^{i}}^{W}=\frac{1}{2}\frac{1}{16\pi^2}$ and $C_{{\rm f}_{r}^{i}}^{G}=\frac{1}{2}\frac{1}{16\pi^2}$ are the anomaly coefficients depending on the gauge group representation of the chiral fermions and four-vector currents $K_{Y}^{\mu}$, $K_{W}^{\mu}$ and $K_{G}^{\mu}$ are given by \cite{h1}
\begin{equation}\label{jbjlqq21}
		\begin{split}
			&
			K_{Y}^{\mu}={g'}^2\epsilon^{\mu\nu\alpha\beta}Y_{\nu}Y_{\alpha\beta},\\&
			K_{W}^{\mu}= g^2\epsilon^{\mu\nu\alpha\beta}\big(W^{a}_{\nu\alpha}W^{a}_{\beta}-\frac{g}{3}f^{abc}W^{a}_{\nu}W^{b}_{\alpha}W^{c}_{\beta}\big),\\&
			K_{G}^{\mu}= g_{\rm s}^2\epsilon^{\mu\nu\alpha\beta}\big(G^{A}_{\nu \alpha}G^{A}_{\beta}-\frac{g_{\rm s}}{3}t^{ABC}G^{A}_{\nu}G^{B}_{\alpha}G^{C}_{\beta}\big).
		\end{split}
\end{equation}
It is important to note that the anomaly equation (\ref{er}) or (\ref{jbjlq2}) does not include perturbative gauge and Yukawa interactions.  It is essential to take these into account when formulating the evolution equations for the asymmetries, which is done in the next section. However, we shall ignore them for most of this section.

Now, Eq.\ (\ref{jbjlq2}) can be expressed in the following form:
 \begin{equation}\label{fakecons}
 \nabla_{\mu} \mathcal{J}_{{\rm f}_{r}^{i}}^{\mu}=0,
 	\end{equation}
where $\mathcal{J}_{{\rm f}_{r}^{i}}^{\mu}=\tilde{j}_{{\rm f}_{r}^{i}}^{\mu}+rN_{\rm c}n_{\rm w}C_{{\rm f}_{r}^{i}}^YK_{Y}^{\mu}-ra_{\rm w}N_{\rm c}C_{{\rm f}_{r}^{i}}^{W}K_{W}^{\mu}+r a_{\rm c}N_{\rm w}C_{{\rm f}_{r}^{i}}^{G}K_{G}^{\mu}$ is our conserved current in the absence of perturbative interactions, and in particular the chirality-flip process. In an expanding Universe with FRW metric, Eq.\ (\ref{fakecons}) can be written out as	
\begin{equation}\label{eq4aq21}
		\partial_{t}\mathcal{J}^{0}+ \frac{1}{R}\vec{\nabla}.\vec{\mathcal{J}}+ 3H\mathcal{J}^{0} =0,
\end{equation}
where the $\mathcal{J}^{0}$ and $\vec{\mathcal{J}}$ are temporal and spatial parts of the current. By utilizing the relation $\dot{s}/s=-3H$ and performing a spatial averaging of Eq.~(\ref{eq4aq21}), we find that the boundary term disappears in the absence of flux, resulting in	
\begin{equation}\label{eq4aq3wsaw1efs}
		\begin{split}
			&\partial_{t}\Big[\frac{N_{\rm c}n_{{\rm f}_{r}^{i}}}{s}+rN_{\rm c}N_{\rm w}\frac{Y_{\rm f}^2}{2}\frac{ n_\mathrm{CS,Y}}{s}-ra_{\rm w}N_{\rm c} \frac{n_\mathrm{\rm CS,w}}{s}+r a_{\rm c}N_{\rm w}\frac{n_\mathrm{CS,s}}{s}\Big]=0.
		\end{split}
\end{equation}
Here, $n_\mathrm{CS,Y}(t)$,  $n_\mathrm{CS,s}(t)$ and $n_\mathrm{\rm CS,w}(t)$ are the Chern-Simons number densities of the ${\rm U}_Y(1)$,  $\rm SU(2)_L$ and $\rm SU(3)$ gauge field configurations, respectively, each of which is given by  \cite{h1,Kuzmin-1985mm,Arnold-987mh,Rubakov-1996vz}
\begin{equation}
		n_{\mathrm{CS}}(t)\equiv\frac{N_{\mathrm{CS}}(t)}{V}=\frac{1}{32\pi^2}\frac{1}{V}\int d^3 x K^0.
\end{equation}
Using the weak and strong sphaleron rates, which are defined as the diffusion constants for topological numbers $N_{\rm CS,w}(t)$ and $N_{\rm CS,s}(t)$, the evolution of the fermionic asymmetry density, excluding the perturbative processes, is obtained as (see Refs.\ \cite{Moore-2010jd,baryon-2,baryon-1,Moore-1997sn,DOnofrio-2012phz,baryon-8,Kuzmin-1985mm})	
	\begin{equation}\label{eq4aq3wsaw1eq}
		\begin{split}
	\partial_{t}\Big[\frac{n_{{\rm f}_{r}^{i}}}{s}\Big]&=\partial_{t}\Big[-\frac{rY_{\rm f}^2N_{\rm w}}{2s}\langle\vec{A}_{Y}\cdot\vec{B}_{Y}\rangle\Big]+ra_{\rm w}\Gamma_{\rm w}\frac{n_{\rm ws}}{2s}+ r a_{\rm c}\Gamma_{\rm s}\frac{n_{\rm ss}}{s},
		\end{split}
	\end{equation} 
where $\Gamma_{\rm w}\simeq25\alpha_{\rm w}^5T$ and $\Gamma_{\rm s}\simeq100\alpha_{\rm s}^5 T$ are the weak and strong sphaleron rates, and the weak and strong sphaleron coefficients, $n_{\rm ws}$ and $n_{\rm ss}$, are defined as \cite{baryon-8,baryon-2}
	\begin{equation}\label{eqss}
		n_{\rm ws}:= \sum_{i}\left(N_{\rm c}(n_{u_L^i}+n_{d_{L}^i})+n_{e_{L}^i}+n_{\nu_{L}^i}\right),
	\end{equation}
	\begin{equation}\label{eqssq}
		n_{\rm ss}:=\frac{1}{N_{\rm c}}\sum_{i}\big(n_{u_{L}^i}+n_{d_{L}^i}-n_{u_{R}^i}-n_{d_{R}^i}\big).
	\end{equation}
	
At high temperatures, rapid $\rm SU(2)_L$ gauge interactions result in equal number densities for different components of a given $\rm SU(2)_L$ multiplet, {\it i.e.}, $n_{e_{L}^i}=n_{\nu_{L}^i}$ and $n_{u_{L}^i}=n_{d_{L}^i}\equiv n_{Q^i}$. As mentioned earlier, We have also assumed that the fast $\rm SU(3)$ color interaction are in equilibrium. Furthermore, due to the flavor mixing in the quark sector, all up or down quarks belonging to different generations with distinct handedness have the same chemical potential; {\it i.e.,}  $n_{u_{R}^i}=n_{u_R}$, $n_{d_{R}^i}=n_{d_R}$ and $n_{Q^i}=n_{Q}$  \cite{86}. Therefore, Eqs.\ (\ref{eqss}) and (\ref{eqssq}) can be simplified to
\begin{equation}\label{eqss1}
	n_{\rm ws}=2\left(9 n_{Q}+n_{e_{L}}+n_{\mu_{L}}+n_{\tau_L}\right),
\end{equation}
\begin{equation}\label{eqssq1}
	n_{\rm ss}=\big(2n_{Q} -n_{u_{R}}-n_{d_{R}}\big).
\end{equation}
Setting $n_{\rm ss}=0$ amounts to having the strong sphaleron processes in absolute equilibrium, which is an excellent approximation in the temperature range under consideration. However, to calculate $\Delta N_{\rm CS,w}$ for the weak sphaleron processes, as will be explained below, we should not set $n_{\rm ws}=0$, even though it would have otherwise been a very good approximation. In App.\ \ref{ame}, we write the AMHD equations by taking into account the CME, and derive the evolution equations for the hypermagnetic field amplitudes. In the next section, we present the complete set of coupled evolution equations for the total baryon asymmetry, the chiral lepton asymmetries and the hypermagnetic field amplitudes, taking into account all perturbative and nonperturbative effects. We then derive the expression for the  global conserved charge, and in particular show the cancellation of the perturbative contributions.

\section{The evolution equations and the conserved charges}\label{4}
	
The evolution of each fermion species is governed by various nonperturbative and perturbative processes. Although the perturbative effects do not appear explicitly in lepton and baryon number violation, they have important effects on their evolution. In particular, the chirality-flip processes play a major role in the presence of the weak sphaleron processes. Indeed, the right-handed fermions get converted to the left-handed ones through the Yukawa interaction, and then the weak sphaleron processes tend to wash them out.
Starting with Eq.\ (\ref{eq4aq3wsaw1eq}), taking in to account the chirality-flip processes, and assuming that the fast perturbative gauge interactions are in equilibrium, we obtain the evolution of the right-handed and the left-handed electron, muon and tau as follows \cite{baryon-1,baryon-8} 
\begin{equation}\label{as1}
	\begin{split}
		&\frac{d\eta_{{e}_{R}}}{dt}=-\frac{g'^{2}}{8\pi^{2}}\frac{d}{dt}\langle\frac{\vec{A}_{Y}\cdot\vec{B}_{Y}}{s}\rangle+\Gamma_{e}\left(\eta_{e_{L}}-\eta_{e_{R}}-\frac{\eta_0}{2}\right),\\&
		\frac{d\eta_{{\mu}_{R}}}{dt}=-\frac{g'^{2}}{8\pi^{2}}\frac{d}{dt}\langle\frac{\vec{A}_{Y}\cdot\vec{B}_{Y}}{s}\rangle+\Gamma_{\mu}\left(\eta_{\mu_{L}}-\eta_{\mu_{R}}-\frac{\eta_0}{2}\right),\\&
		\frac{d\eta_{{\tau}_{R}}}{dt}=-\frac{g'^{2}}{8\pi^{2}}\frac{d}{dt}\langle\frac{\vec{A}_{Y}\cdot\vec{B}_{Y}}{s}\rangle+\Gamma_{\tau}\left(\eta_{\tau_{L}}-\eta_{\tau_{R}}-\frac{\eta_0}{2}\right),\\&
		\frac{d\eta_{{e}_{L}}}{dt}=\frac{d\eta_{{\nu}_{e_L}}}{dt}=\frac{g'^{2}}{32\pi^{2}}\frac{d}{dt}\langle\frac{\vec{A}_{Y}\cdot\vec{B}_{Y}}{s}\rangle-\frac{1}{2}\Gamma_{e}\left(\eta_{e_{L}}-\eta_{e_{R}}-\frac{\eta_0}{2}\right)-\frac{1}{2}\Gamma_{\rm w}\left(9\eta_{Q}+\eta_{e_{L}}+\eta_{\mu_{L}}+\eta_{\tau_L}\right),\\&
		\frac{d\eta_{{\mu}_{L}}}{dt}=\frac{d\eta_{{\nu}_{\mu_L}}}{dt}=\frac{g'^{2}}{32\pi^{2}}\frac{d}{dt}\langle\frac{\vec{A}_{Y}\cdot\vec{B}_{Y}}{s}\rangle-\frac{1}{2}\Gamma_{\mu}\left(\eta_{\mu_{L}}-\eta_{\mu_{R}}-\frac{\eta_0}{2}\right)-\frac{1}{2}\Gamma_{\rm w}\left(9\eta_{Q}+\eta_{e_{L}}+\eta_{\mu_{L}}+\eta_{\tau_L}\right),\\&
		\frac{d\eta_{{\tau}_{L}}}{dt}=\frac{d\eta_{{\nu}_{\tau_L}}}{dt}=\frac{g'^{2}}{32\pi^{2}}\frac{d}{dt}\langle\frac{\vec{A}_{Y}\cdot\vec{B}_{Y}}{s}\rangle-\frac{1}{2}\Gamma_{\tau}\left(\eta_{\tau_{L}}-\eta_{\tau_{R}}-\frac{\eta_0}{2}\right)-\frac{1}{2}\Gamma_{\rm w}\left(9\eta_{Q}+\eta_{e_{L}}+\eta_{\mu_{L}}+\eta_{\tau_L}\right),
		\end{split}
\end{equation}
where $\eta_0$ is the asymmetry of the Higgs field and we have used the relation $\eta=n/s=\mu cT^{2}/6s$ with $c=1$ for fermions and $c=2$ for bosons.  In the above equations, $s=2\pi^{2}g^{*}T^3/45$ is the entropy density, $g^*=106.75$ denotes the effective number of relativistic degrees of freedom, $t_\mathrm{EW}=\left(M_{0}/2T_\mathrm{EW}^{2}\right)$ and $M_{0}=\left(M_\mathrm{Pl}/1.66\sqrt{g^{*}}\right)$, $M_\mathrm{Pl}$ being the Plank mass. Moreover, $\Gamma_i\simeq10^{-2}h_{i}^{2}T/8\pi=\Gamma_{i}^{0}/(\sqrt{x}t_{\rm EW})$ are lepton-Yukawa interactions rate with $\Gamma_{e}^{0}=11.38$, $\Gamma_{\mu}^{0}=4.88\times10^{5} $,  $\Gamma_{\tau}^{0}=1.45\times10^{8}$. Moreover, $\Gamma_{\rm w} \simeq 25 \alpha_W^5 T =\Gamma_{\rm ws}^0 / t_{\rm EW} \sqrt{x}$  is the weak sphaleron rate, where $\alpha_W \simeq 3.17 \times 10^{-2}$,  $\Gamma_{\rm ws}^0 = 2.85 \times 10^9$ and $x=\left(t/t_\mathrm{EW}\right)=\left(T_\mathrm{EW}/T\right)^{2}$, in accordance with the Friedmann law \cite{baryon-8}. In this set of equations, the weak sphaleron coefficient is precisely $n_{\rm ws}$, which, as mentioned above, is allowed to vary. As we shall see in the time evolution problem, it remains very close to zero, indicating equilibrium condition, since its coefficient $\Gamma_{\rm w}$ is extremely large. However, as we shall see below, their product is finite and contributes to $\Delta N_{\rm CS,w}$.

Adding up the chiral quarks asymmetries and using $\eta_{ B}=3(2\eta_{Q}+\eta_{u_R}+\eta_{d_R})=12\eta_Q$, the perturbative quark chirality flip processes cancel out and we obtain the following total baryon asymmetry evolution equation,
\begin{equation}\label{as2}
	\begin{split}
		\frac{d\eta_{ B}}{dt}=-\frac{3g'^{2}}{16\pi^{2} }\frac{d}{dt}\langle\frac{\vec{A}\cdot\vec{B}}{s}\rangle-3\Gamma_{\rm w}\left(9\eta_{Q}+\eta_{e_{L}}+\eta_{\mu_{L}}+\eta_{\tau_L}\right).
	\end{split}
\end{equation} 	
The hypercharge conservation and neutrality of the plasma can be written as \cite{h1}
\begin{equation}
	\eta_Y\equiv -2(\eta_{e_R}+\eta_{\mu_R}+\eta_{\tau_R})-2(\eta_{e_L}+\eta_{\mu_L}+\eta_{\tau_L})+9 (-\frac{2}{3}\eta_{d_R}+\frac{4}{3}\eta_{u_R}+\frac{2}{3}\eta_{Q})+2\eta_{0}=0.
\end{equation}
 Using this condition, the Higgs asymmetry is found as a function of all other chiral leptons and baryon asymmetries,
\begin{equation}
	\eta_{0}=\frac{2}{11}\left(\eta_{e_R}+\eta_{\mu_R}+\eta_{\tau_R}+\eta_{e_L}+\eta_{\mu_L}+\eta_{\tau_L}-\frac{1}{2}\eta_{ B}\right).
\end{equation}
Adding all of the evolution equations for chiral leptons and baryon asymmetries given by Eqs.\ (\ref{as1}), (\ref{as2}), the leptonic chirality flip processes also cancel out and we obtain the conservation law,
\begin{equation}\label{BL12}
	\frac{d}{dt}\left[\eta_{ B+L} +\frac{3g'^{2}}{8\pi^{2} s}\langle\vec{A}_{Y}\cdot\vec{B}_{Y}\rangle +6\int dt \Gamma_{\rm w}\left(\frac{3}{4}\eta_{ B}+\eta_{e_{L}}+\eta_{\mu_{L}}+\eta_{\tau_L}\right)\right]=0, 
\end{equation}
where $\eta_{ B+L}=\eta_{ B}+\eta_{L_e}+\eta_{L_{\mu}}+\eta_{L_{\tau}}$ with $\eta_{L_{\rm f}}=\eta_{{\rm f}_R}+\eta_{{\rm f}_L}+\eta_{\nu_{{\rm f}_L}}$ for ${\rm f}=e,\mu,\tau$. Moreover, the hypermagnetic helicity is $h_B=\langle\vec{A}_{Y}\cdot\vec{B}_{Y}\rangle$. Therefore, in addition to the well known conserved global charge $B-L$, there is another conserved global charge of the electroweak plasma \cite{h1,baryon-2}, which we refer to as the `Matter-Gauge (MG) charge',
\begin{equation}\label{BLwq1}
	\eta_{\rm MG}=\eta_{ B+L} +\eta_{\vec{A}_{Y}\cdot\vec{B}_{Y}}+\eta_{\rm CS,w} ,
\end{equation}
where $\eta_{\vec{A}_{Y}\cdot\vec{B}_{Y}}\equiv\frac{3g'^{2}}{8\pi^{2} s}\langle\vec{A}_{Y}\cdot\vec{B}_{Y}\rangle$ and $\eta_{\rm CS,w}= 6\int dt \Gamma_{\rm w}\left(\frac{3}{4}\eta_{ B}+\eta_{e_{L}}+\eta_{\mu_{L}}+\eta_{\tau_L}\right)$. 
As can be observed in Eq.\ (\ref{BL12}), the chirality-flip processes do not directly appear in the MG charge. However, they have a significant impact on the evolution of $\eta_{ B+L}(x)$ and $\rm N_{\rm CS,w}(x)$.

By considering Eqs.\ (\ref{as1}), (\ref{as2}), and the single mode helical hypermagnetic fields considered in the appendix, with components $B_a$ and $B_d$, and the definition and evolution given by Eqs.~(\ref{eq15})  and (\ref{eq28}), and utilizing the relations $1\,\text{Gauss}\approx 2\times 10^{-20}\,\text{GeV}^2$ and $x=t/t_{\rm EW}=(T_{\rm EW}/T)^2$  all of the coupled evolution equations can be collected and expressed as 
\begin{equation}\label{eq47}
	\begin{split}
		&\frac{d \eta_{e_R}(x)}{dx}=-C_{1}\frac{d}{dx}\left[\left(\bar{B}_{a}^{2}(x)-\bar{B}_{d}^{2}(x)\right)x^2\right]-\frac{\Gamma_{e}^{0}}{\sqrt{x}}\eta_{e,\rm Yuk}(x),\\&
		\frac{d\eta_{\mu_{R}}(x)}{dx}=-C_{1}\frac{d}{dx}\left[\left(\bar{B}_{a}^{2}(x)-\bar{B}_{d}^{2}(x)\right)x^2\right]-\frac{\Gamma_{\mu}^{0}}{\sqrt{x}}\eta_{\mu,\rm Yuk}(x),\\&
		\frac{d\eta_{\tau_{R}}(x)}{dx}=-C_{1}\frac{d}{dx}\left[\left(\bar{B}_{a}^{2}(x)-\bar{B}_{d}^{2}(x)\right)x^2\right]-\frac{\Gamma_{\tau}^{0}}{\sqrt{x}}\eta_{\tau,\rm Yuk}(x),\\&
		\frac{d \eta_{e_L}(x)}{dx}=+\frac{1}{4}C_{1}\frac{d}{dx}\left[\left(\bar{B}_{a}^{2}(x)-\bar{B}_{d}^{2}(x)\right)x^2\right]+\frac{\Gamma_{e}^{0}}{2\sqrt{x}}\eta_{e,\rm Yuk}(x)-\frac{\Gamma_{\rm ws}^{0}}{2\sqrt{x}}\eta_{\rm ws}(x),\\&
		\frac{d\eta_{\mu_{L}}(x)}{dx}=+\frac{1}{4}C_{1}\frac{d}{dx}\left[\left(\bar{B}_{a}^{2}(x)-\bar{B}_{d}^{2}(x)\right)x^2\right]+\frac{\Gamma_{\mu}^{0}}{2\sqrt{x}}\eta_{\mu,\rm Yuk}(x)-\frac{\Gamma_{\rm ws}^{0}}{2\sqrt{x}}\eta_{\rm ws}(x),\\&
		\frac{d\eta_{\tau_{L}}(x)}{dx}=+\frac{1}{4}C_{1}\frac{d}{dx}\left[\left(\bar{B}_{a}^{2}(x)-\bar{B}_{d}^{2}(x)\right)x^2\right]+\frac{\Gamma_{\tau}^{0}}{2\sqrt{x}}\eta_{\tau,\rm Yuk}(x)-\frac{\Gamma_{\rm ws}^{0}}{2\sqrt{x}}\eta_{\rm ws}(x),\\&
		\frac{d\eta_{B}(x)}{dx}=-\frac{3}{2}C_{1}\frac{d}{dx}\left[\left(\bar{B}_{a}^{2}(x)-\bar{B}_{d}^{2}(x)\right)x^2\right]-3\frac{\Gamma_{\rm ws}^{0}}{\sqrt{x}}\eta_{\rm ws}(x),\\&
		\frac{dB_{a}(x)}{dx}=\frac{356k^{\prime\prime}}{\sqrt{x}}\left[-\frac{k^{\prime\prime}}{10^{3}}+C_2\eta_{\rm CM}(x)\right]B_{a}(x)-\frac{B_{a}(x)}{x},\\&
		\frac{dB_{d}(x)}{dx}=\frac{-356k^{\prime\prime}}{\sqrt{x}}\left[-\frac{k^{\prime\prime}}{10^{3}}-C_2\eta_{ \rm CM}(x)\right]B_{d}(x)-\frac{B_{d}(x)}{x},\\
	\end{split}
\end{equation}
where,
\begin{equation}
	\begin{split}
		&\eta_{\rm CM}=\Big[\eta_{e,\rm Yuk}+\eta_{\mu, \rm Yuk}+\eta_{\tau,\rm Yuk}+\frac{1}{4}\eta_{\rm ws}    \Big],	\\&	\eta_{\rm ws}=\left(\frac{3}{4}\eta_{ B}+\eta_{e_{L}}+\eta_{\mu_{L}}+\eta_{\tau_L}\right),\\&
		\eta_{i,\rm Yuk}=\eta_{i_R}-\eta_{i_L}+\frac{1}{2}\eta_{0}\qquad \text{for}\quad  i=e, \mu, \tau.	
	\end{split}
\end{equation}
In the above expressions, we have  defined $k^{\prime\prime}=k/10^{-7}$, $\bar{B}_{j}(x)= B_{j}(x)/10^{20}G$, $C_{1}=\alpha_Y/(5\pi M k^{\prime\prime})$ and $C_2=6\times10^{4}(\alpha_{Y}M/\pi)$, with $M=2\pi^{2}g^{*}/45$ and $\alpha_{Y}={g^{\prime}}^{2}/4\pi$. 
In  Sec.\ \ref{discussion}, the anomalous evolution equations are solved numerically before the EWPT. In particular, the evolution of the weak sphaleron Chern-Simons number, the hypermagnetic helicity and the matter-antimatter asymmetries are obtained.

\section{Numerical Solution}\label{discussion}

\begin{figure}[ht!]
	\centering
	\subfigure[]{\label{figf6}
		\includegraphics[width=.45\textwidth]{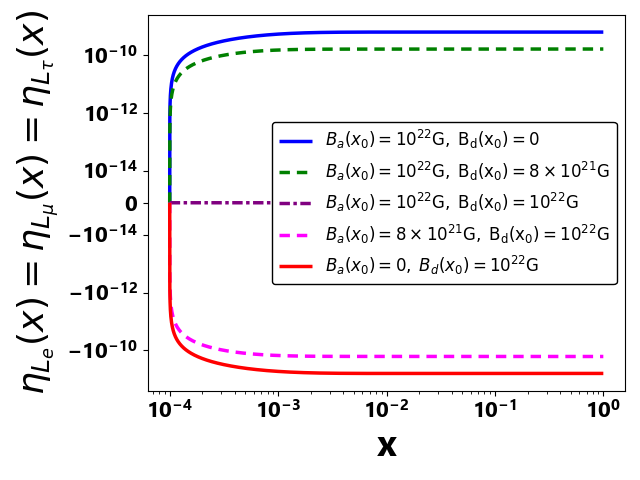}}	
	\hspace{8mm}
	\subfigure[]
	{\label{figf1} 
		\includegraphics[width=.45\textwidth]{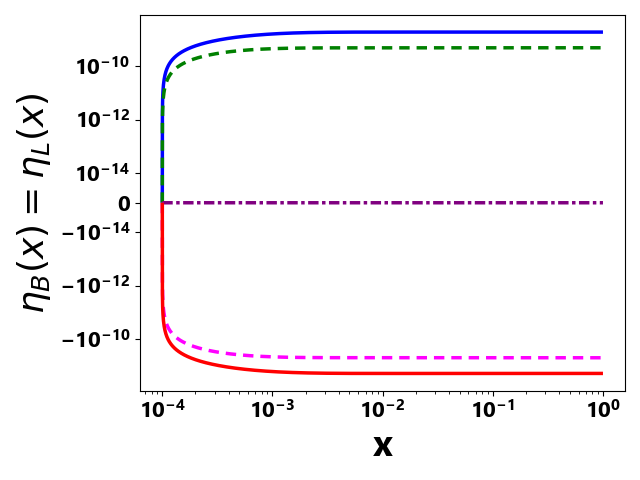}}
	\hspace{8mm}
	\subfigure[]{\label{figf2} 
		\includegraphics[width=.45\textwidth]{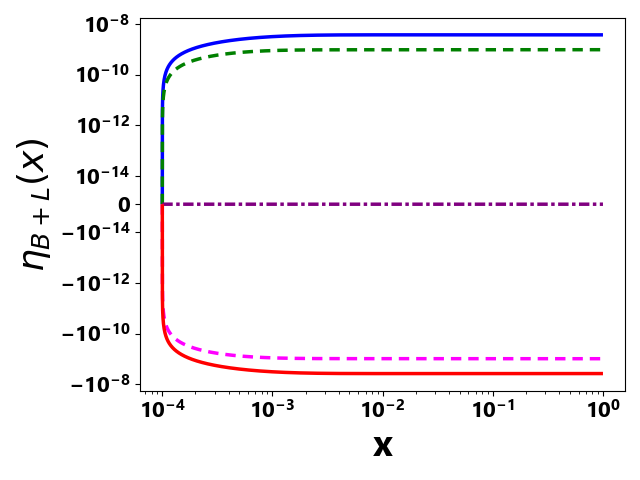}}
	\hspace{8mm}
	\subfigure[]{\label{figf3}
		\includegraphics[width=.45\textwidth]{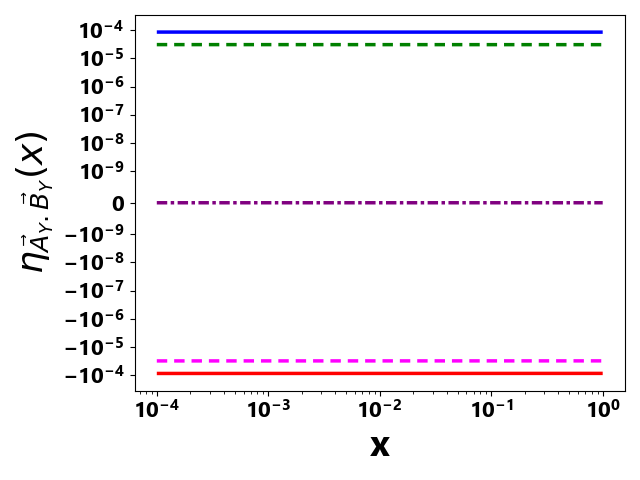}}
	\hspace{8mm}
	\subfigure[]{\label{figf4} 
		\includegraphics[width=.45\textwidth]{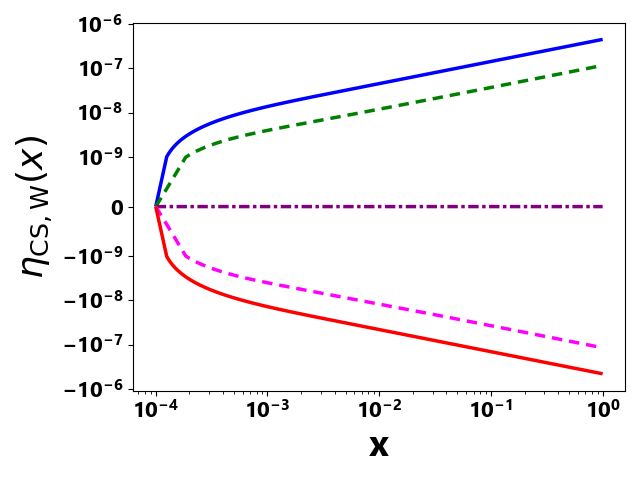}}
	\hspace{8mm}
	\subfigure[]{\label{figf5}
		\includegraphics[width=.45\textwidth]{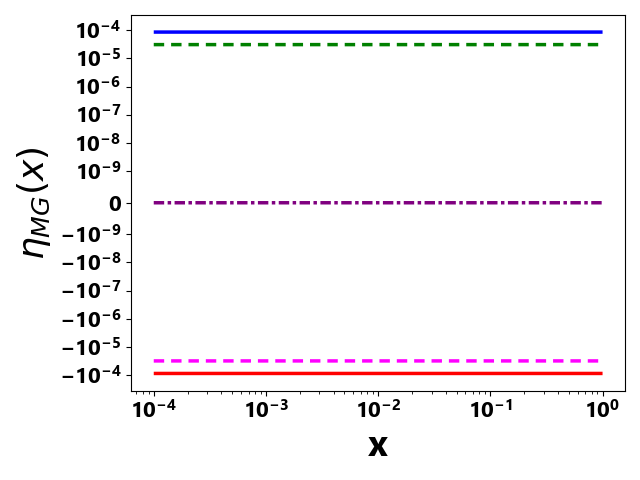}}
	\caption{\footnotesize Time plots of: (a) electron, muon, and tau lepton asymmetry, $\eta_{L_e}(x)=\eta_{L_{\mu}}(x)=\eta_{L_{\tau}}(x)$, (b) baryon and lepton asymmetries $\eta_{ B}(x)=\eta_{L}(x)$, (c) baryon plus lepton asymmetry $\eta_{ B+L}(x)$, (d) hypermagnetic helicity denoted by $\eta_{\vec{A}_Y\cdot\vec{B}_Y}(x)$, (e) Chern-Simons number $\eta_{\rm CS,w}(x)$ , (f) total conserved charge $\eta_{\rm MG}(x)$, for various values of the amplitude of helical components. The initial conditions are:  $k=10^{-8}$, $\eta_{\rm f}^{(0)}=0$. The solid (blued) line is obtained for $B_{a}(x_0)=10^{22} \mbox{G}$ and $ B_{d}(x_0)=0$, the dashed (green) line for $B_{a}(x_0)=10^{22} \mbox{G}$ and $ B_{d}(x_0)=8\times10^{21} \mbox{G}$, the dashed-doted (purple) line for $B_{a}(x_0)=B_{d}(x_0)=10^{22} \mbox{G}$, the dashed (magenta) line for $B_{a}(x_0)=8\times10^{21} \mbox{G}$ and $ B_{d}(x_0)=10^{22} \mbox{G}$, and the solid (red) line for $B_{a}(x_0)=0$ and $ B_{d}(x_0)=10^{22} \mbox{G}$. }
		\label{fig1}
\end{figure}

In this section, we numerically solve the evolution equations derived in Sec.~\ref{4}, in the temperature range $100\, \mbox{GeV}\leq T\leq10\, \mbox{TeV}$, assuming an initial background hypermagnetic field. The initial temperature is chosen to be $T_i = 10\text{ TeV}$, since this is the highest temperature below which all Yukawa interactions are active, and hence serves as a benchmark. A higher initial temperature only extends the early evolution phase-- requiring proportionally stronger initial fields to account for dilution-- while leaving the final asymmetries approximately unchanged. We focus on a scenario where the hypermagnetic field is the sole source for generating the baryon and lepton asymmetries, with all initial asymmetries set to zero. As shown in Eq.~(\ref{bbb}), $B_a$ and $B_{d}$ components have opposite helicities. Consequently, the initial \rm MG conserved charge can be expressed as 	
\begin{align}\label{charget}
		\eta_{\rm MG}(x_0)=\eta_{\vec{A}_{Y}\cdot\vec{B}_{Y}}(x_0)&= \frac{3g'^{2}}{8\pi^{2} s(x_0)}\langle\vec{A}_{Y}(x_0)\cdot\vec{B}_{Y}(x_0)\rangle\notag\\&=\frac{3\alpha_{Y}x_{0}^2}{5\pi^{2} M k^{\prime\prime}}\left[\left(\frac{{B}_{a}(x_0)}{10^{20}\rm G}\right)^2-\left(\frac{{B}_{d}(x_0)}{10^{20}\rm G}\right)^2\right].
\end{align}
For ${B}_{a}(x_0)={B}_{d}(x_0)$, the value of $\eta_{\rm MG}(x_0)$ is zero, and the asymmetries are expected to remain at zero. First, we solve the set of coupled differential equations given in Eq.\ (\ref{eq47}) with the initial conditions  $k=10^{-8}$,  $\eta_{\rm f}^{(0)}=0$, and five different sets of values for the initial hypermagnetic fields ($B_{a}(x_0),B_{d}(x_0)$): \{($10^{22}\mbox{G},0$), ($10^{22}\mbox{G},8\times10^{21}\mbox{G}$), ($10^{22}\mbox{G},10^{22}\mbox{G}$), ($8\times 10^{21}\mbox{G},10^{22}\mbox{G}$), ($0,10^{22}\mbox{G}$)\}, all in units of Gauss. The results are displayed in Fig.\ \ref{fig1}. 

\begin{figure}[ht!]
	\centering
		\includegraphics[width=.55\textwidth]{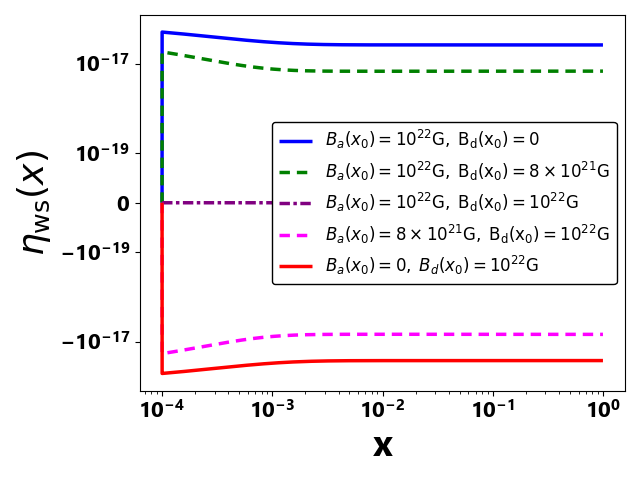}

	\caption{\footnotesize This figure displays the time evolution of the  $\eta_{\rm ws}(x)$  under the same initial conditions as in Figure \ref{fig1}.}
	\label{FFig2}
\end{figure}
 As shown in Figs.\ (\ref{figf6}-\ref{figf2}), the matter-antimatter asymmetries are generated and amplified from zero initial values only for the case $B_{a}(x_0)\neq B_{d}(x_0)$. The source for the generation of all fermion asymmetries is the hypermagnetic helicity. Indeed, in the presence of an initially  strong hypermagnetic field with zero net helicity (see Eq.\ (\ref{charget})), which we shall refer to as non-helical hereafter, the only source term for asymmetry generation is zero, and in back-reaction, the chiral magnetic coefficients which appear with different signs in $B_{a}$ and $B_{d}$ evolution equations remain zero. Subsequently, the hypermagnetic field components $B_{a}$ and $B_{d}$ evolve only through the diffusion and adiabatic expansion terms in  Eq.\ (\ref{eq47}), which are the same for both of them. Therefore, the initially non-helical hypermagnetic field remains non-helical during evolution (see Fig.\ (\ref{figf3})). In contrast, when the initial hypermagnetic field is helical, excess matter or antimatter can be generated from zero initial values, depending on whether $B_{a}>B_{d}$ or $B_{a}<B_{d}$, respectively. Thus, as can be seen in  Figs.\ (\ref{figf6}-\ref{figf2}), within our initial conditions, the maximum values of the excess matter (antimatter) are obtained for the case $ B_{a}(x_0)=10^{22}\mbox{G} $ and
$B_{d}(x_0)=0$ ($B_{a}(x_0)= 0$ and $B_{d}(x_0)=10^{22}\mbox{G}$).

In Fig.\ (\ref{figf4}), we display the time evolution of the $\rm SU(2)_L$ Chern-Simons number $\eta_{\rm CS,w}(x)$ for the aforementioned initial conditions. The results show that in the absence of the generation of the $B+L$ asymmetry, the Chern-Simons number $\eta_{\rm CS,w}(x)$ remains unchanged at its initial value of zero. However, once the asymmetry is generated, $\eta_{\rm CS,w}(x)$ changes correspondingly indicating the action of weak sphaleron processes, which are active in the temperature range of our interest. Moreover, as mentioned previously, in scenarios where the initial helicity is positive (negative), the resulting $\eta_{ B+L}$ is positive (negative). Consequently, the Chern-Simons number undergoes an increase (decrease) from its initial value of zero, leading to a final value of $ \eta_{\rm CS,w}>0$ ($ \eta_{\rm CS,w}<0$). Furthermore, the results show that increasing the initial helicity results in an increase in the $\eta_{ B+L}$ produced, as well as an increase in $\eta_{\rm CS,w}$. Figure \ref{figf3} illustrates the time variation of the hypermagnetic helicity $\eta_{\vec{A}_{Y}\cdot\vec{B}_{Y}}(x)$. While the production of $\eta_{ B+L}$ and $ \eta_{\rm CS,w}$ occur at the expense of $\eta_{\vec{A}_{Y}\cdot\vec{B}_{Y}}(x)$, the magnitude of change is less than $10^{-6}$, which is hardly visible in this figure. It is important to note that an initial $h_B$ decays mostly to $N_{\rm CS,w}$, with only $10^{-3}$ conversion ratio into $B+L$ asymmetry. In Fig.~\ref{figf5} we plot the time variation of the MG charge $\eta_{\rm MG}=\eta_{ B+L} +\eta_{\vec{A}_{Y}\cdot\vec{B}_{Y}}+\eta_{\rm CS,w}$. As is apparent, this charge is conserved, to within the accuracy of our numerical analysis which is $10^{-16}$, for all the cases displayed. 
 
In Fig.~\ref{FFig2}, we present the time evolution of $\eta_{\rm ws}(x)$, the dimensionless coefficient multiplying the weak sphaleron rate $\Gamma_{\rm WS}$ in the evolution equations (Eqs.~(\ref{eq47})). Despite the small amplitude of $\eta_{\rm ws}(x)$, which indicates that these processes stay very close to equilibrium, the integral of its product with the large rate of weak sphaleron processes, $\Gamma_{\rm ws}^{0} \simeq 2.85 \times 10^{9}$, amounts to a magnitude comparable to the change in hypermagnetic helicity  and significantly exceeds the net $B+L$ asymmetry generated in the system. Had we set $\eta_{\rm ws}(x)=0$, which would have set these processes to be in absolute equilibrium, we could not evaluate $\Delta \eta_{\rm CS,w}$.

\begin{figure}[ht!]
	\centering
	\subfigure[]{\label{fig:nB-con}
		\includegraphics[width=.45\textwidth]{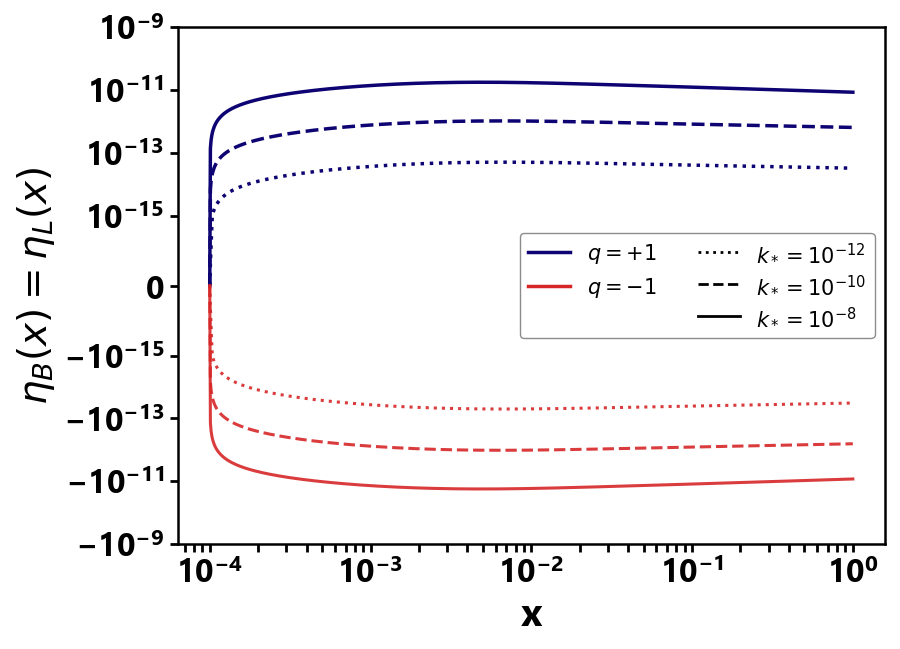}}	
	\hspace{8mm}
	\subfigure[]
	{\label{fig:nBL-con} 
		\includegraphics[width=.45\textwidth]{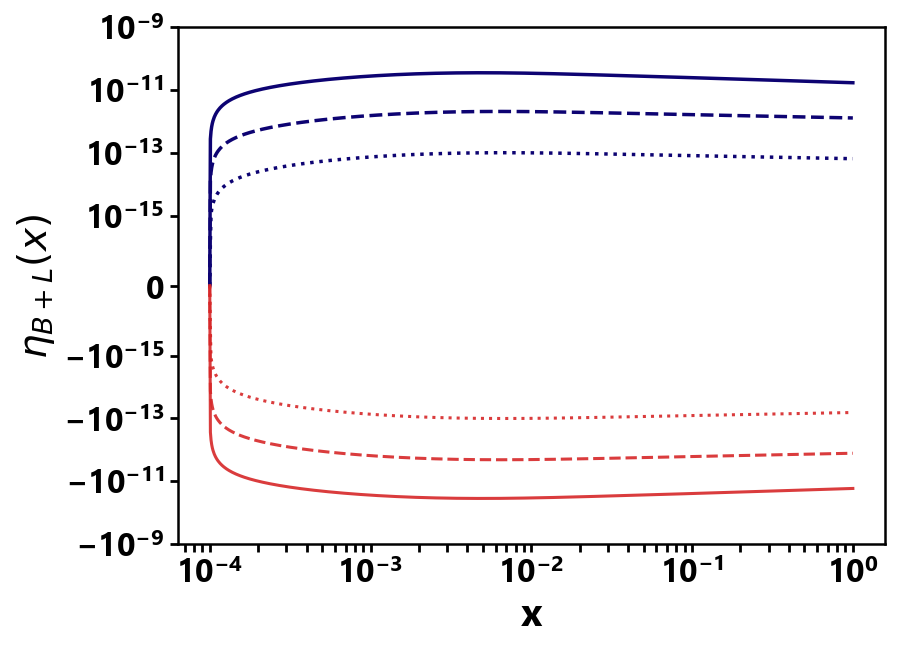}}
	\hspace{8mm}
	\subfigure[]{\label{fig:nAB-con} 
		\includegraphics[width=.45\textwidth]{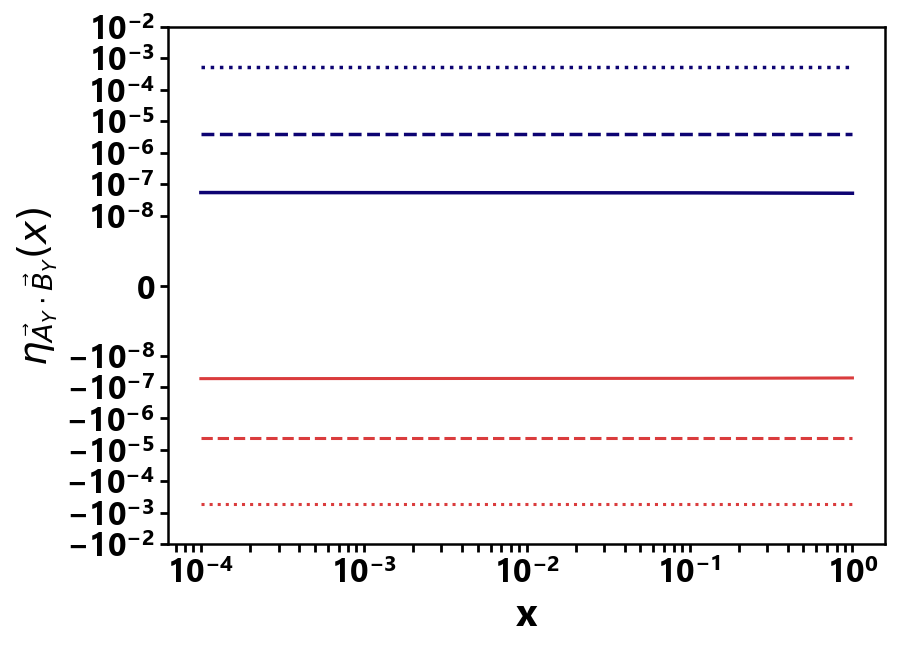}}
	\hspace{8mm}
	\subfigure[]{\label{fig:nCWS-con}
		\includegraphics[width=.45\textwidth]{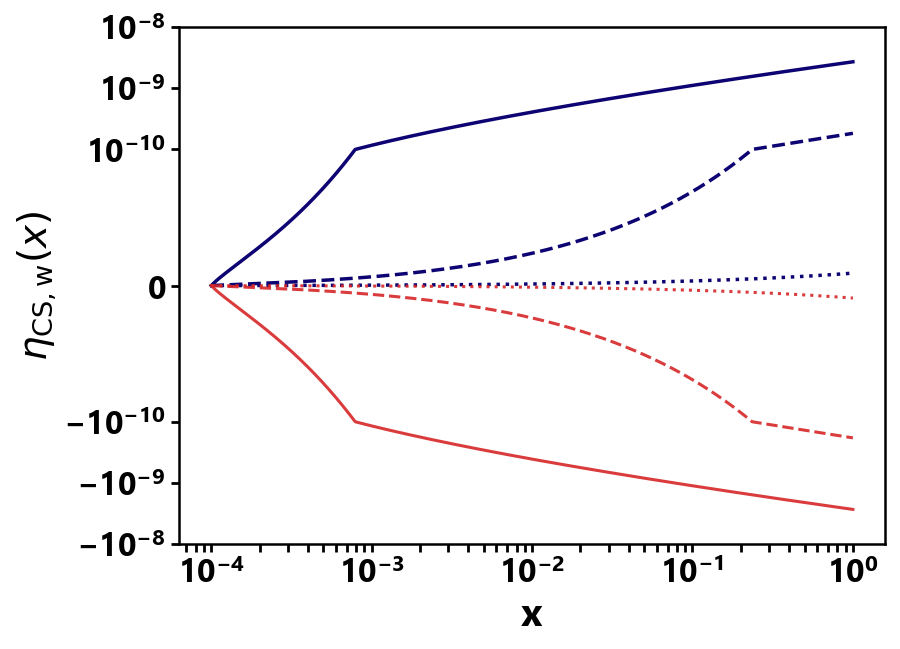}}
	\hspace{8mm}
	\subfigure[]{\label{fig:nWS-con} 
		\includegraphics[width=.45\textwidth]{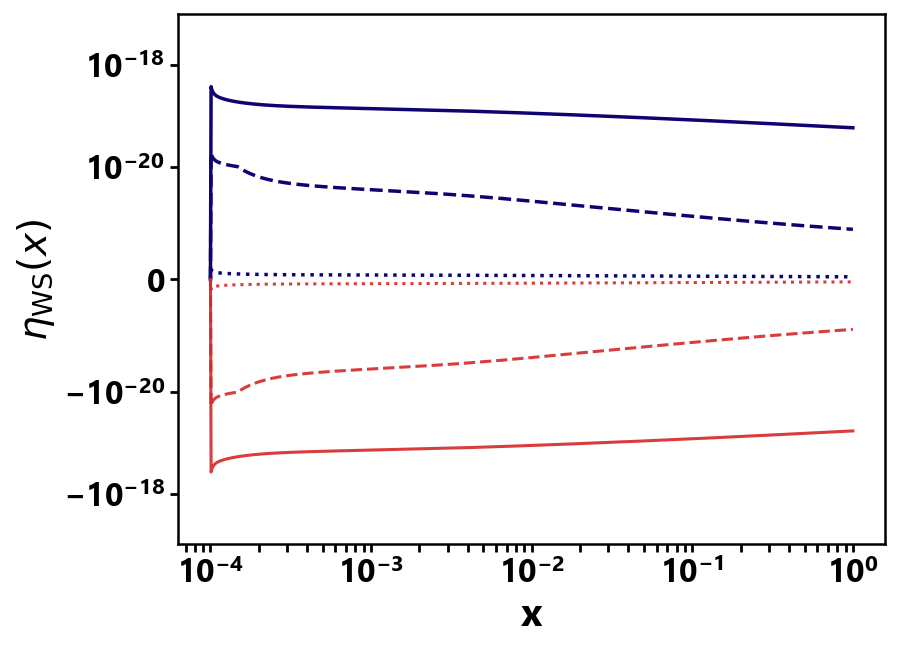}}
	\hspace{8mm}
	\subfigure[]{\label{fig:nMG-con}
		\includegraphics[width=.45\textwidth]{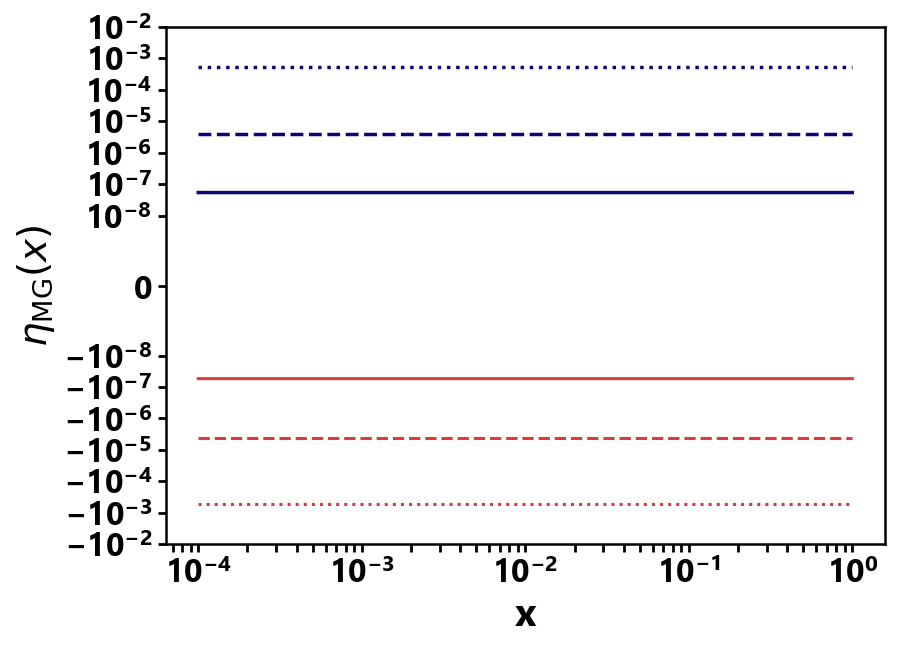}}
	\caption{\footnotesize  Time evolution of: (a) baryon and lepton asymmetries $\eta_B(x) = \eta_L(x)$, (b) baryon plus lepton asymmetry $\eta_{B+L}(x)$, (c) hypermagnetic helicity $\eta_{\vec{A}_Y \cdot \vec{B}_Y}(x)$, (d) Chern--Simons number $\eta_{\rm CS,w}(x)$, (e) sphaleron asymmetry combination $\eta_{\rm ws}(x)$, and (f) total conserved charge $\eta_{\rm MG}(x)$, for continuous hypermagnetic field spectra across various initial hypermagnetic field strengths. Dark blue and red curves correspond to $q = 1$ and $q = -1$, respectively. The solid, dashed, and dotted lines indicate three values of $k_\star$: $10^{-12}$, $10^{-10}$, and $10^{-8}$. The initial conditions are set to $B(x_0) = 10^{22}\,\text{G}$ and $\eta_{\rm f}^{(0)} = 0$.}
	\label{fig3}
\end{figure}

\begin{figure}[ht!]
	\centering
	\subfigure[]{\label{fig:EB}
	\includegraphics[width=.45\textwidth]{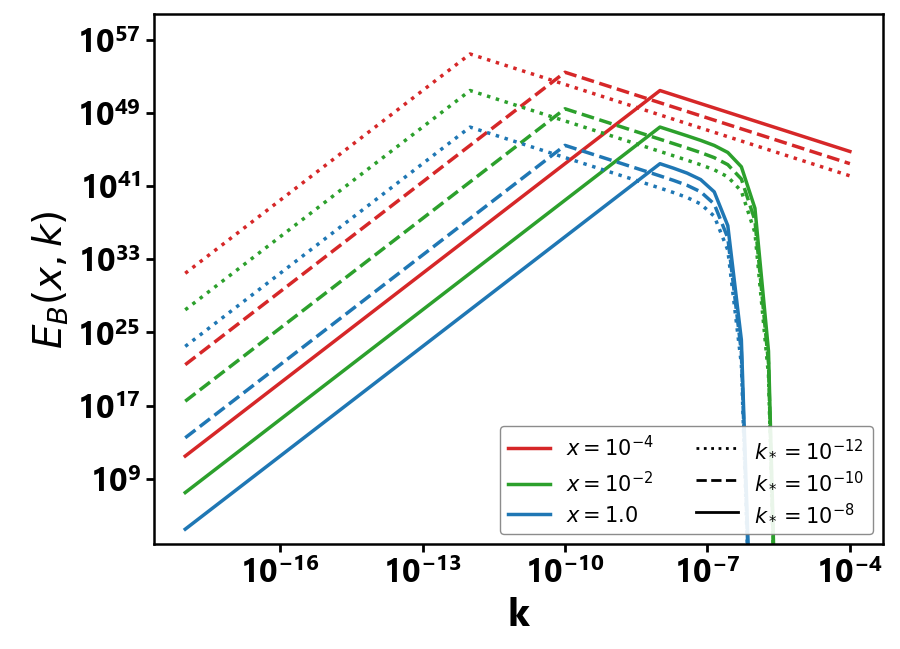}}	
	\hspace{8mm}
	\subfigure[]
	{\label{fig:HB} 
	\includegraphics[width=.45\textwidth]{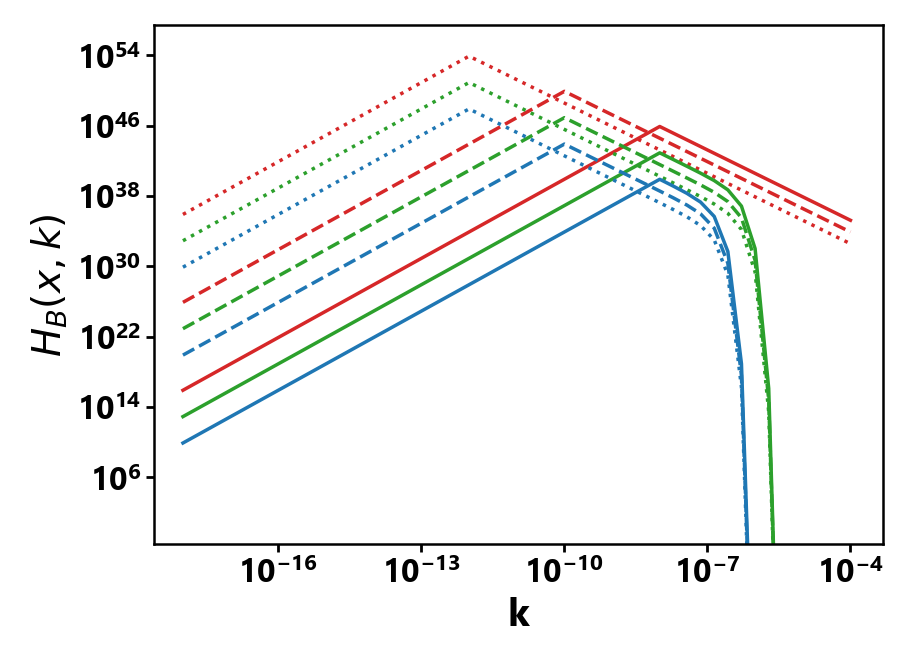}}
		\hspace{8mm}
	\caption{\footnotesize  (a) The hypermagnetic field energy density spectra $E_B(x,k)$ and (b) the helicity density spectra $H_B(x,k)$ at different times. The red, green, and blue curves correspond to $x = 10^{-4}$, $x = 10^{-2}$, and $x = 1$, respectively. The dotted, dashed, and solid lines indicate three values of $k_\star$: $10^{-12}$, $10^{-10}$, and $10^{-8}$. The initial conditions are set to $B(x_0) = 10^{22}\,\text{G}$ and $\eta_{\rm f}^{(0)} = 0$.}
	\label{fig4}
\end{figure}

Finally, we investigate the simultaneous evolution of the hypermagnetic energy spectrum, $E_B(x,k)$, and the hypermagnetic helicity spectrum, $H_B(x,k)$, along with $\eta_{B+L}(x)$, $\eta_{\vec{A}_Y\cdot \vec{B}_Y}(x)$ and  $\eta_{\rm CS,W}(x)$. To achieve this, we adopt a helical basis to extend the monochromatic Chern-Simons wave configurations of the hypermagnetic field to continuous spectra (see Appendix \ref{APP-B}). We fix the normalization by the initial integrated magnetic energy $\int_0^{k_{\max}} dk E_B(x_0,k)=\frac{1}{2}B(x_0)^2$, with  $B(x_0)$ introduced as a convenient measure of an effective magnetic field. We solve the coupled differential equations given in Eq.~(\ref{continues-system}) with the following initial conditions: $B(x_0) = 10^{22}\,\text{G}$ and $\eta_{\rm f}^{(0)} = 0$, for three different values of $k_\star$: $10^{-12}$, $10^{-10}$, and $10^{-8}$, and for two fully helical hypermagnetic fields with positive and negative helicities. The initially fully positive-helicity mode corresponds to $B^{-}(x,k)=0$ with $q=1$, while the fully negative-helicity mode corresponds to $B^{+}(x,k)=0$ with $q=-1$ in Eq.~(\ref{eq:HB-EB}). The transition wave number $k_{\star}$ denotes the scale separating the Batchelor and Kolmogorov regimes in the hypermagnetic energy spectrum. The results are depicted in Figs.~(\ref{fig3}) and (\ref{fig4}). As can be seen from the figures, similar to monochromatic Chern-Simons wave configurations case, in scenarios where the initial helicity is positive (negative), the resulting $\eta_{B+L}$ is positive (negative). Consequently, the Chern-Simons number undergoes an increase (decrease) from its initial value of zero, leading to a final value of $ \eta_{\rm CS,w}>0$ ($ \eta_{\rm CS,w}<0$).  For a continuous spectrum, as shown in Fig \ref{fig:nAB-con}, the variation of the hypermagnetic helicity, $\eta_{\vec{A}_Y \cdot \vec{B}_Y}$, is smaller compared to monochromatic Chern-Simons wave configurations. This is because, for a continuous spectrum, the hypermagnetic energy and helicity are concentrated at $k_{\star}$, where $k_{\star} < 10^{-7}$. Consequently, the diffusion term in the former case is less effective, leading to a smaller variation in $\eta_{\vec{A}_Y \cdot \vec{B}_Y}$ throughout the evolution. Increasing $k_{\star}$ to values larger than $10^{-8}$ yields behavior similar to that shown in Fig.~\ref{figf3} for monochromatic Chern-Simons wave configurations; however, both the baryon asymmetry and the $\rm SU(2)$ Chern-Simons number will decrease for a continuous spectrum.  A critical aspect of the continuous-spectrum model is the sensitivity of the initial hypermagnetic helicity spectrum, $H_B(x_0,k)$, to the transition wave number $k_{\star}$. The transition wave number $k_\star$ determines the peak of the initial Batchelor-Kolmogorov energy spectrum given in Eq.~\eqref{eq:seed_spectrum}, thereby controlling the distribution of magnetic energy and helicity across spatial scales. Increasing the transition wave number $k_{\star}$ shifts the characteristic scale of the initial hypermagnetic spectrum toward shorter wavelengths, while the total initial magnetic energy has been fixed.	Since the helicity spectrum is related to the energy spectrum by $H_B(x_0,k)\propto E_B(x_0,k)/(kT_{EW})$, the integrated helicity is dominated by long-wavelength modes. Consequently, a larger value of $k_{\star}$ reduces the initial hypermagnetic helicity and therefore lowers the initial value of the conserved matter-gauge (MG) charge corresponding to that particular initial condition. On the other hand, the magnetic diffusion rate grows as $k^2$, so the increased contribution of higher-wavenumber modes enhances the early-time helicity dissipation. Through the Abelian chiral anomaly, this faster conversion of hypermagnetic helicity into fermionic charge produces larger baryon and lepton asymmetries, as well as $\rm SU(2)$ Chen-Simons number even though the total initial helicity reservoir is smaller.

	\newpage
\section{Conclusion}\label{x5}
	
In this work we have examined a matter-antimatter asymmetry generation process in the symmetric phase of the early Universe in the temperature range $ 100\, \mbox{GeV}<T<10 \, \mbox{TeV}$, and in the presence of a background hypermagnetic field. We have taken into account the Abelian ${\rm U}(1)_Y$ anomaly, as well as the $\rm SU(2)_L$ and $\rm SU(3)_{\rm c}$ non-Abelian anomalies. The latter is taken into account in the form of a constraint, since it does not contribute to the conserved charge and its sphaleronic processes are extremely fast. We have also taken into account the CME, but not the CVE. Moreover, the perturbative interactions have also been considered, including the fast gauge and Yukawa interactions. The former are taken into account  in the form of constraints, and the latter in the form of the chirality-flip processes for all fermions. We have then calculated the time evolution of the chemical potentials of all fermions and the Higgs, as well as the hypermagnetic field amplitude, resulting in the total matter-antimatter asymmetry $\eta_{ B+L}$ and the hypermagnetic field helicity $\eta_{\vec{A}_{Y}\cdot\vec{B}_{Y}}$. As is well-known, an asymmetry $\eta_{ B+L}$ can be generated, starting from zero initial value, at the expense of the hypermagnetic field helicity $\eta_{\vec{A}_{Y}\cdot\vec{B}_{Y}}$. To find the exact conservation law, we have also calculated the change in the $\rm SU(2)_L$ Chern-Simons number $\eta_{\rm CS,w}$ due to the weak sphaleron processes.
To calculate this, we have first shown that the coefficient of the weak sphaleron processes in the evolution equations, $\eta_{\rm ws}(x)$, remains extremely close to zero, as expected, indicating near equilibrium condition. The expression that we have presented to calculate  $\Delta \eta_{\rm CS,w}$, involves the time integral of  $\eta_{\rm ws}(x)$ multiplied by $\Gamma_{\rm w}$. While the former is very small, the latter is very large, which amounts to a finite value for $\Delta \eta_{\rm CS,w}$.
We have then verified explicitly the conservation of  the Matter-Gauge charge 
$\eta_{\rm MG}=\eta_{ B+L} +\eta_{\vec{A}_{Y}\cdot\vec{B}_{Y}}+\eta_{\rm CS,w}$, in addition to $\eta_{ B-L}$. 
In particular we have shown that, in the scenario that we have studied, an initial $h_B$ decays mostly to $N_{\rm CS,w}$, with only $10^{-3}$ conversion ratio into $B+L$ asymmetry. Furthermore, We have extended the monochromatic Chern-Simons wave configurations to continuous spectra of hypermagnetic field. Comparing Fig. \ref{fig3} with Fig. \ref{fig1}, we note that the qualitative behavior of the generated baryon and lepton asymmetries, the evolution of hypermagnetic helicity, as well as $\rm SU(2)$ Chern-Simons number in the continuous spectra case are similar to the corresponding results for the monochromatic Chern-Simons wave configuration case. However, for the two cases which can be compared quantitatively {\it, i.e.}, $k=10^{-8}$ for the single-mode case and the $k_{\star}=10^{-8}$ for the continuous spectrum, the generated $\eta_{B+L}(x_{\rm EW})$ and $\eta_{\rm CS, w}(x_{\rm EW})$ for the latter is about two orders of magnitude smaller than the former, when their initial energies are equal.\\

{\bf Acknowledgments:} S. A. acknowledges the support of the Iran National Science Foundation (INSF) (grant No.\ 4003903). SA also acknowledges support by the European Union’s Framework Programme for Research and Innovation Horizon 2020 under the Marie SklodowskaCurie grant agreement No 860881-HIDDeN as well as under the Marie Sklodowska-Curie Staff Exchange grant agreement No 101086085-ASYMMETRY.

\appendix\label{App-a}	
\section{Anomalous Maxwell equations in the symmetric phase of the early Universe}\label{ame}
The anomalous Maxwell equations govern the behavior of the electromagnetic fields in the presence of the  CME and CVE.
In an expanding Universe, the Anomalous Maxwell equations for the hypercharge neutral plasma, taking into account the CME, are given as follows \cite{Giovannini-2013oga,Giovannini-2013oga-2}  
\begin{equation}\label{eq1e}
	\frac{1}{R}\vec{\nabla}\cdot\vec{E}_{Y}=\rho_{ \rm total}=0,\qquad\qquad\frac{1}{R}\vec{\nabla}\cdot\vec{B}_{Y}=0,
\end{equation}
\begin{equation}\label{eq2}
	\frac{1}{R}\vec{\nabla}\times\vec{ E}_{Y}+\left(\frac{\partial \vec{B}_{Y}}{\partial t}+2H\vec{B}_{Y}\right)=0,
\end{equation}
\begin{equation}\label{eq3e}
	\begin{split}
		\frac{1}{R}\vec{\nabla}\times\vec{B}_{Y}-\left(\frac{\partial \vec{E}_{Y}}{\partial t}+2H\vec{E}_{Y}\right)=\vec{J}, 
	\end{split}
\end{equation}
\begin{equation}\label{eq2.3q}
	\begin{split}
		\vec{J}&=\vec{J}_{\mathrm{Ohm}}+\vec{J}_{\mathrm{cm}}
		\\&
		=\sigma\vec{E}_{Y}+c_{\mathrm{B}}\vec{B}_{Y}
	\end{split}, 
\end{equation}
The coefficient $c_{\mathrm{B}}$ for massless fermions in the symmetric phase of the early Universe is given as follows \cite{baryon-6,baryon-7,baryon-8,s3,s4,avi1}
\begin{equation}\label{eq26a}
	\begin{split} 
		&c_{\mathrm{B}}(t)=\frac{-g'^{2}}{8\pi^{2}} \sum_{i=1}^{n_{G}}\left[-2\mu_{e_R^i}+\mu_{e_L^i}-\frac{2}{3}\mu_{d_R^i}-\frac{8}{3}\mu_{u_R^i}+\frac{1}{3}\mu_{Q^{i}}\right],		
	\end{split}
\end{equation}
where $n_{G}$ is the number of generations and $\mu_{e_R^i}$ ($\mu_{e_L^i}$), $\mu_{u_R^i}$ ($\mu_{d_R^i}$) and $\mu_{Q^i}$ denote the chemical potential of right-handed (left-handed) charged leptons, right-handed up (down) quarks, and left-hand quarks, respectively.
  
Now we choose the following configurations for our hypermagnetic field \cite{ms1}
\begin{equation}\label{eq15}
	\vec{A}_{Y}(t,z) =A_{a}(t)\hat{a}(z,k)+A_{d}(t)\hat{d}(z,k),
\end{equation}	
where $\hat{a}(z,k)=(\cos kz, -\sin kz,0)$ and  $\hat{d}(z,k)=( \cos kz,\sin kz, 0)$ are common Chern-Simons configurations with positive and negative helicity, respectively. These topologically nontrivial configurations have been used extensively to solve the MHD equations \cite{Giovannini-2013oga-2,Giovannini-2016whv,Giovannini-2013oga}. Using Eq.\ (\ref{eq15})  the hypermagnetic field is obtained as
\begin{equation}\label{bbb1}
	\begin{split}
		\vec{B}_{Y}=\frac{1}{R} \vec{\nabla}\times\vec{A}_{Y}&= \frac{k}{R} A_{a}(t)\hat{a}(z,k)-\frac{k}{R}A_{d}(t)\hat{d}(z,k),\\&
		=B_{a}(t)\hat{a}(z,k)-B_{d}(t)\hat{d}(z,k).
	\end{split}
\end{equation}
In this equation, $B_a$ and $B_{d}$ correspond to the positive and negative helical components of the hypermagnetic field, respectively. Using Eq.\ (\ref{bbb1}), we obtain the hypermagnetic field energy and helicity densities as 
\begin{equation}\label{bbb}
	\begin{split}
		&	\rho_{ B}=\frac{1}{2}\langle\vec{B}_{Y}.\vec{B}_{Y}\rangle= \frac{1}{2}\left( B_{a}(t)^2 +B_{d}(t)^2 \right),\\&
		h_{ B}= \langle\vec{A}_{Y}.\vec{B}_{Y}\rangle= \frac{R}{k}( B_{a}^2(t)-B_{d}^2(t)).
	\end{split}
\end{equation}
When $B_{a}=B_{d}$, the hypermagnetic field becomes completely non-helical. On the other hand, when either $B_{a}\neq 0$ and $B_{d}=0$, or $B_{d}\neq 0$ and $B_{a}=0$, the hypermagnetic field becomes fully helical, with positive or negative helicity, respectively.
Upon neglecting the displacement current in Eq.\ (\ref{eq3e}), we can express the hyperelectric field in terms of the hypermagnetic field components,
\begin{equation}\label{eq14}
	\begin{split}
		\vec{E}_{Y}&=\frac{1}{R\sigma}\vec{\nabla}\times\vec{B}_{Y}-\frac{c_{ B}}{\sigma}\vec{B}_{Y}\\&
		=
		\left[\frac{k^{\prime}}{\sigma}B_{a}(t)-\frac{c_\mathrm{B}}{\sigma}B_{a}(t)\right]\hat{a}(z,k)+
		\left[\frac{k^{\prime}}{\sigma}B_{d}(t)+\frac{c_\mathrm{B}}{\sigma}B_{d}(t)\right]\hat{d}(z,k).
	\end{split}
\end{equation} 
Equations (\ref{eq14}) and (\ref{eq2}) then lead to the evolution equation of the hypermagnetic fields,
\begin{equation}\label{eq28}
	\begin{split}
		\frac{\partial B_{a}(t)}{\partial t}=&\left[-\frac{{k^{\prime}}^{2}}{\sigma}+\frac{{k^{\prime}} c_\mathrm{B}}{\sigma}\right]B_{a}(t)-\frac{B_{a}(t)}{t},\\
		\frac{\partial B_{d}(t)}{\partial t}=&\left[-\frac{{k^{\prime}}^{2}}{\sigma}-\frac{k^{\prime} c_\mathrm{B}}{\sigma}\right] B_{d}(t)-\frac{B_{d}(t)}{t},\\
	\end{split}
\end{equation}	
with $k^{\prime}=k/R=kT$. On the RHS of these equations, the first, second, and third terms correspond to the hypermagnetic diffusion, the chiral magnetic effect, and the hypermagnetic dilution due to the expansion, respectively.

\section{Evolution equations for continuous spectra}\label{APP-B}
In this appendix, we present the evolution equations for the matter-antimatter asymmetries and hypermagnetic fields by extending the Chern-Simons configuration to continuous spectra. We begin by decomposing the vector fields into the divergence-free eigenmodes of the Laplacian operator, \cite{s1}:
\begin{equation}
	\vec{Q}^{\pm}(\vec{k}) = \frac{\vec{e}_1(\vec{k}) \pm i\vec{e}_2(\vec{k})}{\sqrt{2}} \exp(i\vec{k} \cdot \vec{r}),
\end{equation}
where $\vec{e}_3 = \vec{k}/k$ and $(\vec{e}_1, \vec{e}_2, \vec{e}_3)$ constitute a right-handed orthonormal basis. These eigenmodes satisfy
 $\vec\nabla \cdot \vec{Q}^{\pm} = 0$ and $\vec\nabla \times \vec{Q}^{\pm} = \pm k \vec{Q}^{\pm}$, with $\vec{Q}^{\pm*}(-\vec{k}) = \vec{Q}^{\pm}(\vec{k})$.
The hypermagnetic field $\vec{B}_{Y}$ is decomposed as \cite{s1}:

\begin{equation}
	\vec{B}_Y(x, \vec{r}) = \int \frac{d^3k}{(2\pi)^3} \left[ \tilde{B}^+(x, \vec{k}) \vec{Q}^+(\vec{k}) + \tilde{B}^{-}(x, \vec{k}) \vec{Q}^-(\vec{k}) \right].
\end{equation}
where $\tilde{B}^{\pm}(x, \vec{k})$ are Fourier transforms of the magnetic fields with positive and negative helicities, respectively. Assuming statistical isotropy, the ensemble-averaged hypermagnetic energy density is
\begin{equation}\label{E_B}
	\frac{1}{2} \langle |\vec{B}_Y(x, \vec{r})|^2 \rangle \equiv \int d k \, E_B(x, k) = \int \frac{k^2 \, dk}{(2\pi)^2} \left[ |B^+(x, k)|^2 + |B^-(x, k)|^2 \right], 
\end{equation}
while the helicity density is
\begin{equation}\label{H_B}
	\langle \vec{A}_Y \cdot \vec{B}_Y \rangle \equiv \int d k \, H_B(x, k) = \int \frac{k \, dk}{2\pi^2 T} \left[ |B^+(x, k)|^2 - |B^-(x, k)|^2 \right],
\end{equation}
The above expressions follow from the statistically isotropic correlators
\begin{equation}
	\langle \tilde{B}^{\pm*}(x, \vec{k}) \tilde{B}^{\pm}(x, \vec{k'}) \rangle = |B^{\pm}(x, k)|^2 (2\pi)^3 \delta^{(3)}(\vec{k} - \vec{k'}), \quad 
\end{equation}
\begin{equation}
	\langle \tilde{B}^{+*}(x, \vec{k}) \tilde{B}^{-}(x, \vec{k'}) \rangle = \langle  \tilde{B}^{-*}(x, \vec{k})  \tilde{B}^{+}(x, \vec{k'}) \rangle = 0. \quad 
\end{equation}

The hypermagnetic field evolution equation can be decomposed into equations for the modes $\tilde{B}^{\pm}$:
\begin{equation}\label{eq49-1} 
	\begin{split}
	&\frac{d}{dx}\tilde{B}^{+}(x, \vec{k})=\frac{356k^{\prime\prime}}{\sqrt{x}}\left[-\left(\frac{k^{\prime\prime}}{10^{3}}\right)+ C_{2}\eta_{CM}(x)
	\right]\tilde{B}^{+}(x, \vec{k}) -\frac{1}{x}\tilde{B}^{+}(x, \vec{k}),\\&	\frac{d}{dx}\tilde{B}^{-}(x, \vec{k})=\frac{356  k^{\prime\prime}}{\sqrt{x}}\left[-\left(\frac{k^{\prime\prime}}{10^{3}}\right)-C_{2}\eta_{CM}(x) \right]\tilde{B}^{-}(x, \vec{k}) -\frac{1}{x}\tilde{B}^{-}(x, \vec{k}),		
	\end{split}
\end{equation}
 Multiplying the evolution equations for the two helicity modes by $\tilde{B}^{+*}$ and $\tilde{B}^{-*}$, respectively, taking the ensemble average, and using the definitions of the hypermagnetic energy and helicity spectra in Eqs.~(\ref{E_B}) and (\ref{H_B}), we obtain the coupled evolution equations of the hypermagnetic energy and helicity spectra

\begin{equation}\label{eq49-3} 
	\begin{split}
	&\frac{d}{dx}E_B(x,k)=-\frac{0.712{k^{\prime\prime}}^2}{\sqrt{x}}E_B(x,k) + C_{2}\left[\frac{356 k T_{EW} k^{\prime\prime}}{x}
	\right]\eta_{CM}(x)  H_B(x,k)-\frac{2}{x}E_B(x,k),\\&
		\frac{d}{dx}H_B(x,k)=-\frac{0.712{k^{\prime\prime}}^2}{\sqrt{x}}H_B(x,k)+ C_{2}\left[\frac{1424k^{\prime\prime}}{k T_{EW}}
		\right]\eta_{CM}(x)E_B(x,k)-\frac{3}{2x}H_B(x,k).
	\end{split}
\end{equation}
These equations are analogous to the evolution equations given in Refs.~\cite{Dvornikov:2022gkf, baryon-3}, but are expressed here in the physical frame. Now the set of coupled evolution equations becomes:

\begin{equation}\label{continues-system}
	\begin{split}
		&\frac{d \eta_{e_R}(x)}{dx}=-C \frac{d}{dx}\left[x^{3/2} \int_{k_{\text{min}}}^{k_{\text{max}}} dk \,  H_B(x,k)\right]-\frac{\Gamma_{e}^{0}}{\sqrt{x}}\eta_{e,\rm Yuk}(x),\\&
		\frac{d\eta_{\mu_{R}}(x)}{dx}=-C \frac{d}{dx}\left[x^{3/2} \int_{k_{\text{min}}}^{k_{\text{max}}} dk \,  H_B(x,k)\right]-\frac{\Gamma_{\mu}^{0}}{\sqrt{x}}\eta_{\mu,\rm Yuk}(x),\\&
		\frac{d\eta_{\tau_{R}}(x)}{dx}=-C \frac{d}{dx}\left[x^{3/2} \int_{k_{\text{min}}}^{k_{\text{max}}} dk \,  H_B(x,k)\right]-\frac{\Gamma_{\tau}^{0}}{\sqrt{x}}\eta_{\tau,\rm Yuk}(x),\\&
		\frac{d \eta_{e_L}(x)}{dx}=+\frac{1}{4}C \frac{d}{dx}\left[x^{3/2} \int_{k_{\text{min}}}^{k_{\text{max}}} dk \,  H_B(x,k)\right]+\frac{\Gamma_{e}^{0}}{2\sqrt{x}}\eta_{e,\rm Yuk}(x)-\frac{\Gamma_{\rm ws}^{0}}{2\sqrt{x}}\eta_{\rm ws}(x),\\&
		\frac{d\eta_{\mu_{L}}(x)}{dx}=+\frac{1}{4}C \frac{d}{dx}\left[x^{3/2} \int_{k_{\text{min}}}^{k_{\text{max}}} dk \,  H_B(x,k)\right]+\frac{\Gamma_{\mu}^{0}}{2\sqrt{x}}\eta_{\mu,\rm Yuk}(x)-\frac{\Gamma_{\rm ws}^{0}}{2\sqrt{x}}\eta_{\rm ws}(x),\\&
		\frac{d\eta_{\tau_{L}}(x)}{dx}=+\frac{1}{4}C \frac{d}{dx}\left[x^{3/2} \int_{k_{\text{min}}}^{k_{\text{max}}} dk \,  H_B(x,k)\right]+\frac{\Gamma_{\tau}^{0}}{2\sqrt{x}}\eta_{\tau,\rm Yuk}(x)-\frac{\Gamma_{\rm ws}^{0}}{2\sqrt{x}}\eta_{\rm ws}(x),\\&
		\frac{d\eta_{ B}(x)}{dx}=-\frac{3}{2}C \frac{d}{dx}\left[x^{3/2} \int_{k_{\text{min}}}^{k_{\text{max}}} dk \,  H_B(x,k)\right]-3\frac{\Gamma_{\rm ws}^{0}}{\sqrt{x}}\eta_{\rm ws}(x),\\&
		\frac{d}{dx}E_B(x,k)=-\frac{0.712{k^{\prime\prime}}^2}{\sqrt{x}}E_B(x,k) + C_{2}\left[\frac{356 k T_{EW} k^{\prime\prime}}{x}
		\right]\eta_{CM}(x)  H_B(x,k)-\frac{2}{x}E_B(x,k),\\&
		\frac{d}{dx}H_B(x,k)=-\frac{0.712{k^{\prime\prime}}^2}{\sqrt{x}}H_B(x,k)+ C_{2}\left[\frac{1424k^{\prime\prime}}{k T_{EW}}
		\right]\eta_{CM}(x)E_B(x,k)-\frac{3}{2x}H_B(x,k),
	\end{split}
\end{equation}
where $C = 10^{-36} \alpha_Y/(2\pi^3 (106.75/45)\times0.5^{1.5})$.

For the numerical solution of coupled evolution equations (\ref{continues-system}), we adopt a composite Batchelor--Kolmogorov spectrum for the initial hypermagnetic field \cite{Dvornikov:2022gkf, baryon-3}. The initial hypermagnetic energy spectrum at $x=x_0$ is taken in the form
\begin{equation}
	E_B(x_0,k)=
	\begin{cases}
		C_B k^{n_B}, & 0<k<k_{\star},\\
		
		C_K k^{n_K}, & k_{\star}<k<k_{\max},
	\end{cases}
	\label{eq:seed_spectrum}
\end{equation}
where $n_B=4$ corresponds to the Batchelor spectrum for small momenta, while $n_K=-5/3$ corresponds to the Kolmogorov spectrum for large momenta. Here we choose $k_{\rm min}=10^{-18}$, $k_{\rm max}=10^{-4}$, and three different values for $k_{\star}=\{10^{-12}, 10^{-10}, 10^{-8} \}$ \cite{Dvornikov:2022gkf}.
The continuity of the initial spectrum at the transition momentum requires

\begin{equation}
	C_B k_\star^{n_B}=C_K k_\star^{n_K},
	\label{eq}
\end{equation}
 which fixes the relative normalization of the two spectral branches. The overall normalization can be determined by defining and specifying an initial effective hypermagnetic field strength $B_0$ through

\begin{equation}\label{CKCB}
	\begin{split}
		\frac{1}{2}B(x_0)^2 &=\int_0^{k_{\max}} dk E_B(x_0,k),
		 \\&= C_B \int_0^{k_\star} k^{n_B} \, dk + C_K \int_{k_\star}^{k_{\rm max}}k^{n_K} \, dk , 
	\end{split}
\end{equation}
 Eqs.~(\ref{eq}) and (\ref{CKCB}) uniquely determine the constants $C_B$ and $C_K$. The resulting initial spectrum can therefore be written as \cite{Dvornikov:2022gkf}
\begin{equation}
	E_{B}(x_0,k) =\frac{(1 + n_K) }{2} B(x_0)^2 \left[ \frac{n_K - n_B}{1 + n_B} + \frac{ k_{\rm max}^{1 + n_K}}{k_\star^{1 + n_K}} \right]^{-1} \times
	\begin{cases}
		\frac{k^{n_B}}{k_\star^{1 + n_B}}, & 0  < k < k_\star, \\
		\frac{k^{n_K}}{k_\star^{1 + n_K}}, & k_\star < k < k_{\rm max}.
	\end{cases}
\end{equation}

The corresponding initial helicity spectrum is chosen as

\begin{equation}\label{eq:HB-EB}
	H_B(x_0,k)=\frac{2q \sqrt{x_0}}{kT_{EW}}E_B(x_0,k),
\end{equation}

where $-1\le q\equiv \frac{|\tilde{B}^{+}|^2- |\tilde{B}^{-}|^2}{|\tilde{B}^{+}|^2+|\tilde{B}^{-}|^2} \le 1$ is a parameter that determines the initial helicity fraction of the hypermagnetic field. The choice $q=1$ ($q=-1$) corresponds to a maximally positive (negative) helical initial field, while $q=0$ corresponds to a non-helical initial field.

By adding the evolution equations for the matter fields, {\it i.e.,} the first seven equations contained in  Eq.~(\ref{continues-system}), and integrating with respect to $x$, the conserved quantity $\eta_{\rm MG}$ for the continuous spectrum system is:
\begin{equation}
	\frac{d}{dx} \left[ \eta_{B+L}(x) + 3 C \, x^{3/2} \int_{k_{\text{min}}}^{k_{\text{max}}} dk \, H_B(x,k) + 6 \int_{x_0}^{x} \frac{\Gamma_{\rm ws}^0}{\sqrt{x'}} \eta_{\rm ws}(x') \, dx' \right] = 0.
\end{equation}
Therefore, the conserved Matter-Gauge (MG) charge $\eta_{\rm MG}$ is:
\begin{equation}\label{eq:conserved2}
	\eta_{\rm MG} = \eta_{B+L}(x) + \eta_{\vec{A}_Y \cdot \vec{B}_Y}(x) + \eta_{\rm CS, w}(x) = \text{const.},
\end{equation}
where the terms correspond to the total baryon plus lepton asymmetry, the  hypermagnetic helicity contribution, and the weak sphaleron Chern-Simons contribution, respectively.

Now, we show that the monochromatic Chern-Simons basis given by Eq.\ (\ref{eq15}) is a special case of the continuous helicity basis. Consider a plane wave propagating along the $z$ axis with wave vector $\vec{k}=k\hat{z}$. In this case, the right-handed orthonormal basis is 
	\begin{equation}
		\vec{e}_1=\hat{x},\qquad \vec{e}_2=\hat{y},\qquad \vec{e}_3=\frac{\vec{k}}{k}=\hat{z},
	\end{equation}
	and the corresponding helicity eigenmodes are
	\begin{equation}
		\vec{Q}^{\pm}(k,\hat{z}) =\frac{\hat{x}\pm i\hat{y}}{\sqrt{2}}e^{ikz},
	\end{equation}
	which satisfy the curl eigenvalue equation$ \nabla\times\vec{Q}^{\pm} =\pm k \vec{Q}^{\pm}$. The monochromatic Chern-Simons basis vectors $\hat{a}(z,k)$ and $\hat{d}(z,k) $ can be expressed as
	\begin{align}
		\hat{a}(z,k)&=
		\frac{1}{\sqrt{2}}
		\left[
		\vec{Q}^{+}(k\hat{z})
		+
		\vec{Q}^{+*}(k\hat{z})
		\right],\\
		\hat{d}(z,k)
		&=
		\frac{1}{\sqrt{2}}
		\left[
		\vec{Q}^{-}(k\hat{z})
		+
		\vec{Q}^{-*}(k\hat{z})
		\right].
	\end{align}
	Hence, $\hat{a}$ and $\hat{d}$ are the real combinations of the positive- and negative-helicity Fourier eigenmodes, respectively. They satisfy $ \nabla\times\hat{a}=k\hat{a},
	$ and $ \nabla\times\hat{d}=-k\hat{d}$, showing that they represent pure helicity states. The monochromatic formulation is recovered from the continuous description by concentrating the magnetic power at a single wave number,
	\begin{equation}
		|B^{+}(k)|^{2}=\frac{2\pi^2}{k^2} B_{a}^{2}\delta(k-k_{0}),
		\qquad
		|B^{-}(k)|^{2}=\frac{2\pi^2}{k^2} B_{d}^{2}\delta(k-k_{0}),
	\end{equation}
	for which the continuous definitions of the hypermagnetic energy and helicity reduce exactly to the monochromatic expressions,
	\begin{equation}
		\int dk E_{B}(k)=\frac12(B_{a}^{2}+B_{d}^{2}),
		\qquad
		\int dk H_{B}(k)=\frac{R}{k_{0}}(B_{a}^{2}-B_{d}^{2}).
	\end{equation}
	Therefore, the monochromatic Chern--Simons description adopted in this work is simply the specific single mode realization of the general continuous helicity decomposition.

	\newpage





\small
	\begin{thebibliography}{99}
		
		\bibitem{baryon3}
		G. Steigman, Primordial Nucleosynthesis: The Predicted and Observed Abundances and Their Consequences, PoS NICXI (2010) 001, \href{http://arxiv.org/abs/1008.4765}{[arXiv:1008.4765 [astro-ph.CO]]}.
		
		\bibitem{WMAP-2010qai}
		E.~Komatsu \textit{et al.} [WMAP],``Seven-Year Wilkinson Microwave Anisotropy Probe (WMAP) Observations: Cosmological Interpretation,''
		Astrophys. J. Suppl. \textbf{192}, 18 (2011)
		[DOI:10.1088/0067-0049/192/2/18]
		\href{http://arxiv.org/abs/1001.4538}{[arXiv:1001.4538 [astro-ph.CO]]}.
		
		\bibitem{Sakharov1}
		A. D. Sakharov, Violation of CP Invariance, C Asymmetry, and Baryon
		Asymmetry of the Universe, Pisma Zh. Eksp. Teor. Fiz. 5 (1967) 32
		[JETP Lett. 5 (1967) 24] [Sov. Phys. Usp. 34 (1991) 392] [Usp. Fiz.
		Nauk 161 (1991) 61], [DOI: 10.1070/PU1991v034n05ABEH002497].	
		
		\bibitem{1}
		L. M. Widrow, Origin of galactic and extragalactic magnetic fields, Rev. Mod. Phys.  \textbf{74}, 775 (2002), \href{http://arxiv.org/abs/astro-ph/0207240}{[arXiv:astro-ph/0207240]}.
		\bibitem{2}
		P. P. Kronberg, Extragalactic magnetic fields, Rep. Prog. Phys.\textbf{57}, 325 (1994), [DOI: 10.1088/0034-4885/57/4/001] .		
		
		\bibitem{magnetic1}
		M. L. Bernet, F. Miniati, S. J. Lilly, P. P. Kronberg and M. Dessauges-Zavadsky, Strong magnetic fields in normal galaxies at high redshift, Nature 454 (2008) 302, \href{http://arxiv.org/abs/0807.3347}{[arXiv:0807.3347 [astro-ph]]}.
		
		\bibitem{magnetic2}
		F. Tavecchio, G. Ghisellini, L. Foschini, G. Bonnoli, G. Ghirlanda and P. Coppi, The intergalactic magnetic field constrained by Fermi/Large Area Telescope observations of the TeV blazar 1ES 0229+200, Mon. Not. Roy. Astron. Soc. 406 (2010) L70, \href{http://arxiv.org/abs/1004.1329}{[arXiv:1004.1329 [astro-ph.CO]]}.
		\bibitem{magnetic3}
		S. ’i. Ando and A. Kusenko, Evidence for gamma-ray halos around active galactic nuclei and the first measurement of intergalactic magnetic fields, Astrophys. J. 722 (2010) L39, \href{http://arxiv.org/abs/1005.1924}{[arXiv:1005.1924 [astro-ph.HE]]}.
		
		\bibitem{magnetic4}
		A. Neronov and I. Vovk, Evidence for Strong Extragalactic Magnetic Fields from Fermi Observations of TeV Blazars, Science 328 (2010) 73, \href{http://arxiv.org/abs/1006.3504}{[arXiv:1006.3504 [astro-ph.HE]]}.
		\bibitem{magnetic5}
		M. Sydorenkoa, O. Tomalakb, and Y. Shtanova, Magnetic fields and chiral asymmetry in the early hot Universe, JCAP \textbf{10}, (2016), 018, \href{http://arxiv.org/abs/1607.04845}{[arXiv:1007.3891 [astro-ph.CO]]} 
		
		
		
		\bibitem{Neronov:2010bi}
		A.~Neronov, D.~V.~Semikoz, P.~G.~Tinyakov and I.~I.~Tkachev,
		``No evidence for gamma-ray halos around active galactic nuclei resulting from intergalactic magnetic fields,''
		Astron. Astrophys. \textbf{526}, A90 (2011)
		doi:10.1051/0004-6361/201015892
		\href{http://arxiv.org/abs/1006.0164}{[arXiv:1006.0164 [astro-ph.HE]]}.
		
		
		
		\bibitem{Fermi-LAT:2013mql}
		M.~Ackermann \textit{et al.} [Fermi-LAT],
		``Determination of the Point-Spread Function for the Fermi Large Area Telescope from On-orbit Data and Limits on Pair Halos of Active Galactic Nuclei,''
		Astrophys. J. \textbf{765}, no.1, 54 (2013)
		doi:10.1088/0004-637X/765/1/54
		\href{http://arxiv.org/abs/1309.5416}{[arXiv:1309.5416 [astro-ph.HE]]}.
		
		\bibitem{AlvesBatista:2021sln}
		R.~Alves Batista and A.~Saveliev,``The Gamma-ray Window to Intergalactic Magnetism,''
		Universe \textbf{7}, no.7, 223 (2021)
		doi:10.3390/universe7070223
		\href{http://arxiv.org/abs/1309.5416}{2105.12020}{[arXiv:2105.12020 [astro-ph.HE]]}.
		
		\bibitem{Durrer-2013pga}
		R.~Durrer and A.~Neronov,``Cosmological Magnetic Fields: Their Generation, Evolution and Observation,''
		Astron. Astrophys. Rev. \textbf{21}, 62 (2013)
		[DOI:10.1007/s00159-013-0062-7]
		\href{http://arxiv.org/abs/1303.7121}{[arXiv:1303.7121 [astro-ph.CO]]}.
		
		\bibitem{Naoz-2013wla}
		S.~Naoz and R.~Narayan,``Generation of Primordial Magnetic Fields on Linear Over-density Scales,''
		Phys. Rev. Lett. \textbf{111}, 051303 (2013)
		[DOI:10.1103/PhysRevLett.111.051303]
		\href{http://arxiv.org/abs/1304.5792}{[arXiv:1304.5792 [astro-ph.CO]]}.
		
		\bibitem{mf1}
		T. Vachaspati, Magnetic fields from cosmological phase transitions, 
		Phys.Lett. B \textbf{265}, 258261 (1991), [DOI: 10.1016/0370-2693(91)90051-Q].
		
		\bibitem{mf2}
		K. Enqvist and P. Olesen, On primordial magnetic fields of electroweak origin, 
		Phys. Lett. B \textbf{319}, 178 (1993), \href{http://arxiv.org/abs/hep-ph/9308270}{[arXiv:hep-ph/9308270]}.
		
		\bibitem{mf4}
		J. M. Cornwall, Speculations on primordial magnetic helicity, 
		Phys.Rev. D \textbf{56}, 6146 (1997), \href{http://arxiv.org/abs/hep-th/9704022}{[ hep-th/9704022]}.	
		\bibitem{mf5}
		S. M. Carroll, G. B. Field, and R. Jackiw, Limits on a Lorentz and Parity Violating Modification of Electrodynamics,
		Phys. Rev. D \textbf{41}, 1231 (1990), [DOI: 10.1103/PhysRevD.41.1231].	
		\bibitem{mf6}
		M. M. Anber, E. Sabancilar, Hypermagnetic Fields and Baryon Asymmetry from Pseudoscalar Inflation, 
		Phys.Rev. D \textbf{92} (2015) no.10, 101501,\href{http://arxiv.org/abs/1507.00744}{[arXiv:1507.00744]}. 
		\bibitem{mf7}
		R. M. Kulsrud, R. Cen, J. P. Ostriker, and D. Ryu, The Protogalactic Origin for Cosmic Magnetic Fields, 
		Astrophys.J. \textbf{480}, 481 (91997), \href{http://arxiv.org/abs/astro-ph/9607141}{[	arXiv:astro-ph/9607141]}.
		
		\bibitem{Taylor-2011bn}
		A.~M.~Taylor, I.~Vovk and A.~Neronov,``Extragalactic magnetic fields constraints from simultaneous GeV-TeV observations of blazars,''
		Astron. Astrophys. \textbf{529}, A144 (2011)
		[DOI:10.1051/0004-6361/201116441]
		\href{http://arxiv.org/abs/1101.0932}{[arXiv:1101.0932 [astro-ph.HE]]}.
		
		\bibitem{huan}
		H. Huan, T. Weisgarber, T. Arlen, S. P. Wakely, A New Model for Gamma-Ray Cascades in Extragalactic Magnetic Fields,  (2011), Astrophys. J. L \textbf{735} L28, \href{http://arxiv.org/abs/1106.1218}{[arXiv:1106.1218 [astro-ph.HE]}.	
		\bibitem{Vovk}
		I. Vovk, A. M. Taylor, D. Semikoz, A. Neronov, Fermi/LAT observations of 1ES 0229+200: implications for extragalactic magnetic fields and background light, (2012),  Astrophys. J. \textbf{747} L14,  \href{http://arxiv.org/abs/1112.2534}{arXiv:1112.2534 [astro-ph.CO]}.
		
		\bibitem{Kandus-2010nw}
		A.~Kandus, K.~E.~Kunze and C.~G.~Tsagas,``Primordial magnetogenesis,''
		Phys. Rept. \textbf{505}, 1-58 (2011)
		[DOI:10.1016/j.physrep.2011.03.001]
		\href{http://arxiv.org/abs/astro-ph/1007.3891}{[arXiv:1007.3891 [astro-ph.CO]]}.
		
		\bibitem{a1}
		S. L. Adler, Axial-vector vertex in spinor electrodynamics, Phys. Rev. \textbf{177}, 2426 (1969), [DOI:10.1103/PhysRev.177.2426].
		\bibitem{a2}
		J. S . Bell and R. Jackiw, Nuovo Cimento A \textbf{60}, 47 (1969).
		\bibitem{a3}
		G. ’t Hooft, “Symmetry Breaking Through Bell-Jackiw Anomalies,” Phys. Rev. Lett. \textbf{37} (1976), [DOI: 10.1103/PhysRevLett.37.8].
		
		\bibitem{Vilenkin:1978hb}
		A.~Vilenkin,``Parity Violating Currents in Thermal Radiation,'' Phys. Lett. B \textbf{80}, 150-152 (1978) [DOI:10.1016/0370-2693(78)90330-1].	
		\bibitem{av1}
		A. Vilenkin, Macroscopic Parity Violating Effects: Neutrino Fluxes From Rotating Black Holes And In Rotating Thermal Radiation,
		Phys. Rev. D \textbf{20}, 1807 (1979), [DOI: 10.1103/PhysRevD.20.1807].
		\bibitem{avi1}
		A. Vilenkin, Equilibrium Parity Violating Current In A Magnetic Field, 
		Phys. Rev. D \textbf{22}, 3080 (1980), [DOI: 10.1103/PhysRevD.22.3080].
		
		\bibitem{kharz}	
		D. E. Kharzeev, Topology, magnetic field, and strongly interacting matter,
		\href{http://arxiv.org/abs/1501.01336}{[ arXiv:1501.01336 [hep-ph]]}.
		
		\bibitem{Giovannini-1997eg}
		M.~Giovannini and M.~E.~Shaposhnikov,``Primordial hypermagnetic fields and triangle anomaly,''Phys. Rev. D \textbf{57}, 2186-2206 (1998)	[DOI:10.1103/PhysRevD.57.2186]
		\href{http://arxiv.org/abs/hep-ph/9710234}{[arXiv:hep-ph/9710234 [hep-ph]]}.	
		\bibitem{Giovannini-2016whv}
		M.~Giovannini, ``Anomalous magnetohydrodynamics in the extreme relativistic domain,''
		Phys. Rev. D \textbf{94}, no.8, 081301 (2016)
		[DOI:10.1103/PhysRevD.94.081301]
		\href{http://arxiv.org/abs/1606.08205}{[arXiv:1606.08205 [hep-th]]}.
		
		\bibitem{Giovannini-2013oga}
		M.~Giovannini,``Anomalous Magnetohydrodynamics,''	Phys. Rev. D \textbf{88}, 063536 (2013) [DOI:10.1103/PhysRevD.88.063536]
		\href{http://arxiv.org/abs/1307.2454 }{[arXiv:1307.2454 [hep-th]]}.
		\bibitem{s2}
		S. Abbaslu, S. Rostam Zadeh and S. S. Gousheh, Contribution of the chiral vortical effect to the evolution of the hypermagnetic field and the matter-antimatter asymmetry in the early Universe, \href{http://arxiv.org/abs/1908.10105}{[arXiv:1908.10105 [hep-ph]]}.
		
		\bibitem{s3}
		S.~Abbaslu, S.~Rostam Zadeh, M.~Mehraeen and S.~S.~Gousheh, ``The generation of matter-antimatter asymmetries and hypermagnetic fields by the chiral vortical effect of transient fluctuations,''
		Eur. Phys. J. C \textbf{81}, no.6, 500 (2021)
		[DOI:10.1140/epjc/s10052-021-09272-9]
		\href{http://arxiv.org/abs/2001.03499}{[arXiv:2001.03499 [hep-ph]]}.
		
		\bibitem{s4}
		S.~Abbaslu, S.~R.~Zadeh, A.~Rezaei and S.~S.~Gousheh,
		``Effects of nonhelical component of hypermagnetic field on the evolution of the matter-antimatter asymmetry, vorticity, and hypermagnetic field,'' Phys. Rev. D \textbf{104}, no.5, 056028 (2021) [DOI:10.1103/PhysRevD.104.056028]
		\href{http://arxiv.org/abs/2104.05013}{[arXiv:2104.05013 [hep-ph]]}.
		
		
		\bibitem{baryon-5}
		K. Kamada and A. J. Long, evolution of the baryon asymmetry through the electroweak crossover in the presence of a helical magnetic field, Phys. Rev. D \textbf{94}, 123509 (2016), \href{http://arxiv.org/abs/1610.03074}{ arXiv:1610.03074 [hep-ph]}.
		
		\bibitem{Hamada:2025cwu}
		Y.~Hamada, K.~Mukaida and F.~Uchida,
		Symmetries of hot SM, magnetic flux baryogenesis from helicity decay,
		JHEP \textbf{01}, 040 (2026)
		doi:10.1007/JHEP01(2026)040
		\href{http://arxiv.org/abs/2507.01576}{[arXiv:2507.01576 [hep-ph]]}.
		
		\bibitem{Boyer:2025jno}
		T.~Boyer and A.~Neronov,
		Baryon asymmetry constraints on magnetic field from the Electroweak epoch,'
		\href{http://arxiv.org/abs/2504.07937}{[arXiv:2504.07937 [astro-ph.CO]]}.
		
		\bibitem{Hamada:2025ooy}
		Y.~Hamada, K.~Mukaida and F.~Uchida,
		Revisiting constraints on magnetogenesis from baryon asymmetry,'
		Phys. Rev. D \textbf{113}, no.8, 8 (2026)
		doi:10.1103/45lx-vp2h
		\href{http://arxiv.org/abs/2509.23858}{[arXiv:2509.23858 [hep-ph]]}.
		
			\bibitem{Jackiw-1976pf}
		R.~Jackiw and C.~Rebbi,``Vacuum Periodicity in a Yang-Mills Quantum Theory,''
		Phys. Rev. Lett. \textbf{37}, 172-175 (1976) [DOI:10.1103/PhysRevLett.37.172]
		
		\bibitem{Callan-1976je}
		C.~G.~Callan, Jr., R.~F.~Dashen and D.~J.~Gross,``The Structure of the Gauge Theory Vacuum,'' Phys. Lett. B \textbf{63}, 334-340 (1976) [DOI:10.1016/0370-2693(76)90277-X ]
		
		
		\bibitem{Manton-1983nd}
		N.~S.~Manton,``Topology in the Weinberg-Salam Theory,''
		Phys. Rev. D \textbf{28}, 2019 (1983)
		[DOI:10.1103/PhysRevD.28.2019]
		\bibitem{Klinkhamer-1984di}
		F.~R.~Klinkhamer and N.~S.~Manton,``A Saddle Point Solution in the Weinberg-Salam Theory,''Phys. Rev. D \textbf{30}, 2212 (1984)
		[DOI:10.1103/PhysRevD.30.2212]
		
			\bibitem{Kuzmin-1985mm}
		V.~A.~Kuzmin, V.~A.~Rubakov and M.~E.~Shaposhnikov,``On the Anomalous Electroweak Baryon Number Nonconservation in the Early Universe,''
		Phys. Lett. B \textbf{155}, 36 (1985) [DOI:10.1016/0370-2693(85)91028-7]
		
		
		\bibitem{Arnold-987mh}
		P.~B.~Arnold and L.~D.~McLerran,``Sphalerons, Small Fluctuations and Baryon Number Violation in Electroweak Theory,''
		Phys. Rev. D \textbf{36}, 581 (1987)
		[DOI:10.1103/PhysRevD.36.581]
		
		\bibitem{Rubakov-1996vz}
		V.~A.~Rubakov and M.~E.~Shaposhnikov,``Electroweak baryon number nonconservation in the early universe and in high-energy collisions,'' Usp. Fiz. Nau\textbf{166}, 493-537 (1996) [DOI:10.1070/PU1996v039n05ABEH000145]
		\href{http://arxiv.org/abs/hep-ph/9603208}{[arXiv:hep-ph/9603208 [hep-ph]]}.	
		
		\bibitem{Moore-1}
		G.D. Moore, Computing the strong sphaleron rate, Phys. Lett. B \textbf{412} (1997) 359, \href{http://arxiv.org/abs/hep-ph/9705248}{arXiv:[hep-ph/9705248]}.
		
		\bibitem{Moore-2010jd}
		G.~D.~Moore and M.~Tassler,``The Sphaleron Rate in SU(N) Gauge Theory,''
		JHEP \textbf{02}, 105 (2011) [DOI:10.1007/JHEP02(2011)105]
		\href{http://arxiv.org/abs/1011.1167}{[arXiv:1011.1167 [hep-ph]]}.
			\bibitem{ms1}
			M. Giovannini, Primordial Hypermagnetic Knots, Phys. Rev. D\textbf{61}, 063004 (2000), \href{http://arxiv.org/abs/hep-ph/9905358}{[arXiv:hep-ph/9905358]}.	
			
			\bibitem{baryon-4}
			T. Fujita and K. Kamada, Large-scale magnetic fields can explain the baryon asymmetry of the Universe, Phys. Rev. D \textbf{93}, 083520 (2016), \href{http://arxiv.org/abs/1602.02109v2}{arXiv:1602.02109v2 [hep-ph] }.
			
			\bibitem{baryon-1}
			M. Dvornikov and V. B. Semikoz, Lepton asymmetry growth in the symmetric phase of an electroweak plasma with hypermagnetic fields versus its washing out by
			sphalerons, Phys. Rev. D \textbf{87}, 025023 (2013), \href{http://arxiv.org/abs/1212.1416}{[arXiv:1212.1416 [astro-ph.CO]]}.
			\bibitem{Dvornikov:2022gkf}
            M.~Dvornikov,
            ``Impact of hypermagnetic fields on relic gravitational waves, neutrino oscillations and baryon asymmetry,''
            Int. J. Mod. Phys. D \textbf{32} (2023) no.02, 2250141
            doi:10.1142/S0218271822501413 \href{http://arxiv.org/abs/2203.00530}{[arXiv:2203.00530 [astro-ph.CO]]}.
			\bibitem{baryon-2}
			K. Kamada and A. J. Long, Baryogenesis from decaying magnetic helicity, Phys. Rev. D \textbf{94}, 063501 (2016), \href{http://arxiv.org/abs/1606.08891}{[ arXiv:1606.08891 [astro-ph.CO]}.
			

			
			\bibitem{baryon-3}
			V. B. Semikoz, A. Yu. Smirnov, and D. D. Sokoloff, Generation of hypermagnetic helicity and leptogenesis in the early Universe, Phys. Rev. D \textbf{93}, 103003 (2016), \href{http://arxiv.org/abs/1604.02273}{[arXiv:1604.02273 [hep-ph]]}.
			
			
			
			
			
			\bibitem{baryon-6}
			S. Rostam Zadeh and S. S. Gousheh, 	
			Contributions to the $U_{Y}(1)$ Chern-Simons term and the evolution of fermionic asymmetries and hypermagnetic fields, Phys. Rev. D \textbf{94}, 056013 (2016), \href{http://arxiv.org/abs/1512.01942}{[arXiv:1512.01942 [hep-ph]]}.
			\bibitem{baryon-7}
			S. Rostam Zadeh and S. S. Gousheh, Effects of the $U_{Y} (1)$ Chern-Simons term and its baryonic contribution on matter asymmetries and hypermagnetic fields, Phys. Rev. D \textbf{95}, 056001 (2017), \href{http://arxiv.org/abs/1607.00650}{[arXiv:1607.00650 [hep-ph]]}.
			\bibitem{baryon-8}
			S. Rostam Zadeh and S. S. Gousheh, A Minimal System Including Weak Sphalerons for Investigating the Evolution of Matter Asymmetries and Hypermagnetic Fields, Phys. Rev. D \textbf{99}, 096009, (2019), \href{http://arxiv.org/abs/1812.10092}{[arXiv:1812.10092 [hep-ph]]}.
		
		\bibitem{h1}
		A. J. Long, E. Sabancilar, and T. Vachaspati, Leptogenesis  and primordial magnetic fields, J. Cosmol. Astropart. Phys. 02 (2014) 036, \href{http://arxiv.org/abs/1309.2315}{[arXiv:1309.2315 [astro-ph.CO]]}. 
		
		\bibitem{Baym-1997gq}
		G.~Baym and H.~Heiselberg,``The Electrical conductivity in the early universe,''
		Phys. Rev. D \textbf{56}, 5254-5259 (1997)
		[DOI:10.1103/PhysRevD.56.5254]
		\href{http://arxiv.org/abs/astro-ph/9704214}{arXiv:astro-ph/9704214[astro-ph]}. 
		
		
		\bibitem{Arnold-2000dr}
		P.~B.~Arnold, G.~D.~Moore and L.~G.~Yaffe, ``Transport coefficients in high temperature gauge theories. 1. Leading log results,'' JHEP \textbf{11}, 001 (2000)
		[DOI:10.1088/1126-6708/2000/11/001]
		\href{http://arxiv.org/abs/hep-ph/0010177}{[arXiv:hep-ph/0010177 [hep-ph]]}.
		
					\bibitem{Banerjee:2004df}
		R.~Banerjee and K.~Jedamzik,
		Phys. Rev. D \textbf{70}, 123003 (2004)
		doi:10.1103/PhysRevD.70.123003
		[arXiv:astro-ph/0410032 [astro-ph]].
		
		
		\bibitem{Brandenburg:2017neh}
		A.~Brandenburg, T.~Kahniashvili, S.~Mandal, A.~Roper Pol, A.~G.~Tevzadze and T.~Vachaspati,
		Phys. Rev. D \textbf{96}, no.12, 123528 (2017)
		doi:10.1103/PhysRevD.96.123528
		[arXiv:1711.03804 [astro-ph.CO]].
		
		
		\bibitem{RoperPol:2025lgc}
		A.~Roper Pol and A.~S.~Midiri,
		JCAP \textbf{06}, 070 (2026)
		doi:10.1088/1475-7516/2026/06/070
		[arXiv:2501.05732 [gr-qc]].
		
		
		\bibitem{Moore-1997sn}
		G.~D.~Moore, C.~r.~Hu and B.~Muller,``Chern-Simons number diffusion with hard thermal loops,'' Phys. Rev. D \textbf{58}, 045001 (1998)
		[DOI: 10.1103/PhysRevD.58.045001]
		\href{http://arxiv.org/abs/hep-ph/9710436}{[arXiv:hep-ph/9710436 [hep-ph]]}.	
		
		
		\bibitem{DOnofrio-2012phz}
		M.~D'Onofrio, K.~Rummukainen and A.~Tranberg,``The Sphaleron Rate through the Electroweak Cross-over,''
		JHEP \textbf{08}, 123 (2012) [DOI:10.1007/JHEP08(2012)123]
		\href{http://arxiv.org/abs/1207.0685}{[arXiv:1207.0685 [hep-ph]]}.
		\bibitem{86}
		J. M. Cline, K. Kainulainen and K. A. Olive, Protecting
		the Primordial Baryon Asymmetry From Erasure by Sphalerons, Phys.
		Rev. D \textbf{49} (1994) 6394  \href{http://arxiv.org/abs/hep-ph/9401208}{[hep-ph/9401208]}.
		
		\bibitem{Abbaslu:2025ylq}
		S.~Abbaslu, A.~Rezaei, S.~Rostam Zadeh and S.~S.~Gousheh,``The generation of baryon asymmetry and hypermagnetic field by the chiral vortical effect in the presence of sphalerons,''Nucl. Phys. B \textbf{1015}, 116895 (2025)
		doi:10.1016/j.nuclphysb.2025.116895	
			\bibitem{Giovannini-2013oga-2}
		M. Giovannini, Spectrum of anomalous magnetohydrodynamics, Phys. Rev. D \textbf{93}, 103518 (2016), \href{http://arxiv.org/abs/1509.02126}{[arXiv:1509.02126 [hep-th]]}.
		
		\bibitem{s1}
		H. Tashiro, T. Vachaspati, and A. Vilenkin, Chiral effects and cosmic magnetic fields, Phys. Rev. D \textbf{86}, 105033 (2012),  \href{http://arxiv.org/abs/1206.5549}{[arXiv:1206.5549 [astro-ph.CO]]}.	

		
		
		
		
		
		
		
		
		
		
		
		
		
		
		
		
		
		
		
		
		
		
		
		
		
	
	
		
		
		

		
		




	







		
	
	\end{thebibliography}
\end{document}